\documentclass[prb,floatfix,aps,twocolumn,showpacs,superscriptaddress]{revtex4-1}
\usepackage[dvips]{graphics}
\usepackage[english]{babel}
\usepackage{graphicx}
\usepackage{amssymb}
\usepackage{epstopdf}
\usepackage{latexsym}
\usepackage{amsmath}
\usepackage{MnSymbol}
\usepackage{bbold}
\usepackage{mathtools}
\usepackage{color}
\usepackage{cancel}
\usepackage{soul}
\usepackage{array}
\usepackage{multirow}
\usepackage{makecell} 
\usepackage{xcolor}
\usepackage{tikz}
\usepackage{wasysym}

\soulregister\ref7
\soulregister\eqref7
\soulregister\cite7

\begin{document}

\title{
 Entanglement-entropy analysis of critical and topological quantum phases\\ in a frustrated spin-1/2 Heisenberg ladder}
\author{Azadeh Ghannadan}
\affiliation{Institute of Physics, Slovak Academy of Sciences, D\'{u}bravsk\'{a} cesta 9, 845 11, Bratislava, Slovakia}
\author{Hana Vargov\'{a}}
\email[Corresponding author: ]{hcencar@saske.sk}
\affiliation{Institute of Experimental Physics, Slovak Academy of Sciences, Watsonova 47, Košice, 040 01, Slovakia}
\author{Jozef Genzor}
\affiliation{Institute of Physics, Slovak Academy of Sciences, D\'{u}bravsk\'{a} cesta 9, 845 11, Bratislava, Slovakia}
\author{Roman Kr\v{c}m\'{a}r}
\affiliation{Institute of Physics, Slovak Academy of Sciences, D\'{u}bravsk\'{a} cesta 9, 845 11, Bratislava, Slovakia}
\author{Andrej Gendiar}
\email[Corresponding author: ]{andrej.gendiar@savba.sk}
\affiliation{Institute of Physics, Slovak Academy of Sciences, D\'{u}bravsk\'{a} cesta 9, 845 11, Bratislava, Slovakia}

\date{\today}

\begin{abstract}
We investigate the ground-state phase diagram of a frustrated spin-1/2 Heisenberg ladder in the transverse magnetic field with an anisotropic inter-rung exchange coupling $\alpha$. This bond-anisotropic parameter continuously interpolates among the Ising-like limit $\alpha=0$, the isotropic point $\alpha=1$, and the XY-dominated regime $\alpha \gg 1$. Using density-matrix renormalization group calculations within a matrix-product-state framework, we analyze bipartite entanglement, magnetization, and spin correlations to characterize the emergent quantum phases.
We identify six distinct ground-state phases, including rung-singlet, Haldane-like, Tomonaga–Luttinger liquid, canted Ising-ordered, XY-polarized, and ordered ferromagnetic states. While magnetization and local correlations provide a first insight into phase classifications, the finite-size scaling of the entanglement measures offers a more sensitive and unified diagnostics of both gapped and gapless regimes. In the gapless regime, we extract the central charge $c \simeq 1$, confirming Tomonaga–Luttinger liquid behavior, while at the transition between the canted Ising and ferromagnetic phases, we observe $c \simeq 1/2$, consistent with the Ising universality. Finally, we find that both the Haldane-like behavior and the extended critical Tomonaga–Luttinger liquid regime are strongly confined to the vicinity of the isotropic point $\alpha = 1$.
\end{abstract}

\maketitle
\section{Introduction}

Many remarkable discoveries at the end of the $20$th century became the cornerstones of new research directions over the following three decades, particularly in nanoscale physics. The extraordinary quantum properties of two-dimensional graphene reported in $2004$~\cite{Novoselov}, the topological-insulator behavior observed in $2$D HgTe/CdTe quantum wells in $2007$~\cite{Fu, Konig}, and the realization of Bose–Einstein condensates in optical lattices -- ideal low-dimensional quantum systems -- in $2002$~\cite{Giorgini}, among others, marked the beginning of a new era of quantum and nanotechnologies. Consequently, the deeper understanding of low-dimensional physics has become increasingly important.

A special class of low-dimensional quantum systems is the $n$-leg spin-$1/2$ Heisenberg ladder, defined as $n$ parallel chains of ions coupled through "perpendicular rungs". This system is interesting for several reasons. For example, it exhibits a notable even–odd dichotomy: for an even number of chains, the spin-$1/2$ Heisenberg ladder possesses a spin gap in its energy spectrum, accompanied by an exponential decay of the pair spin–spin correlation function, which is characteristic of spin-liquid behavior. In contrast, for odd $n$, gapless excitations and a power-law decay of the spin–spin correlation function are observed, reflecting the Luttinger-liquid phase~\cite{White}. It was later shown that both types of behavior can coexist in the case $n=2$ when an external magnetic field is applied. At a critical field $h_{c_1}$, the spin gap closes, and the Luttinger liquid state with gapless excitations becomes dominant before the full polarization is reached at another critical field $h_{c_2}$~\cite{Chitra, Giamarchi}.

Additional geometric frustration arising from the diagonal interaction $J_{\times}$  between the nearest-neighboring rungs in the two-leg spin-$1/2$ Heisenberg ladder generates further quantum effects, which have been extensively discussed in several interesting studies~\cite{Weihong, Mila, Honecker, Fouet, Almeida, Hu26, Routh}. In the regime of dominant intra-leg interactions ($J_{\rm leg}=J_{\times}>J_{\rm rung}$), the ground-state phase diagram as a function of magnetic field closely resembles that of the non-frustrated two-leg ladder: the spontaneously gapped Haldane phase evolves into a gapless Luttinger-liquid state, which persists until full polarization is reached~\cite{Mila, Honecker}. Both associated phase transitions at the two critical magnetic fields are of second order~\cite{Honecker}. In contrast, in the opposite interaction limit, the system exhibits behavior characteristic of the Ising universality class. The corresponding magnetization curve as a function of magnetic field displays a staircase structure, including an additional plateau at $m=1/2$. As reported by Honecker et al.~\cite{Honecker}, the $m=1/2$ plateau is associated with a long-range-ordered (LRO) pattern in which a singlet state on one rung alternates with a triplet state on the neighboring rung. Moreover, LRO consisting of singlets on every rung has been confirmed as the ground state at small magnetic fields~\cite{Honecker}. Conversely, in this interaction regime, both phase transitions are of first order~\cite{Honecker}. The richness of the resulting quantum-critical behavior is another key reason for the continued scientific interest in this model~\cite{Hung, Weihong, Wang, Nakamura, Fath, Fu21}. Notably, additional forms of quantum criticality have been reported, including spinon and magnon condensations at the edges of (half-)integer magnetization plateaus~\cite{Fouet}, as well as the presence of Kosterlitz–Thouless and bi-critical points~\cite{Almeida}.

Last but not least, the renewed interest in ladder systems is strongly motivated by the wide experimental realization of ladder compounds, including various cuprates, vanadates, rare-earth oxides, the polyradical BIP-BNO compound, and many other materials~\cite{Chaboussant, Chaboussant2, Shiramura, Chaboussant3, Dagotto99, Korotin, Katoh, Watson, Landee, Matsumoto, Oosawa, Batchelor, Zhou, Manabe}.

In this work, we investigate the frustrated spin-1/2 ladder across Ising-like, isotropic Heisenberg, and XY-dominated exchange regimes using the density matrix renormalization group (DMRG) method. Particular attention is devoted to the role of exchange anisotropy in controlling the interplay between frustration, quantum fluctuations, and transverse-field polarization, which gives rise to competing gapped, critical, and symmetry-protected quantum phases. To continuously connect these distinct interaction regimes, we introduce a bond-anisotropy parameter $0 \leq \alpha \leq 2$ acting simultaneously on the intra-leg coupling $J_{\rm leg}$ and the diagonal interaction $J_{\times}$. The limiting cases $\alpha=0$, $\alpha=1$, and $\alpha=2$ correspond respectively to the Ising-like, fully isotropic Heisenberg, and strongly XY-dominated exchange limits. This formulation reveals a rich phase structure comprising gapped ordered states, a Tomonaga–Luttinger liquid regime, and Haldane-like phases with nontrivial entanglement properties. Our analysis demonstrates that both quantum criticality and Haldane-like behavior are strongly tied to the vicinity of the isotropic point, highlighting the pronounced sensitivity of low-dimensional frustrated quantum states to exchange anisotropy. While the gapless Tomonaga–Luttinger liquid is rapidly suppressed by even weak deviations from isotropy, the Haldane-like regime remains comparatively more robust due to its stability at lower transverse fields. Moreover, we show that bipartite entanglement, complemented by finite-size scaling analysis, provides a particularly powerful framework for resolving competing quantum phases and their critical behavior. This approach allows us to distinguish critical and conventionally ordered quantum phase transitions on an equal basis, uncovering universal correlation properties beyond the reach of local magnetic observables alone.

The density matrix renormalization group (DMRG) is a numerical method for calculating ground-state properties of a one-dimensional quantum many-body system~\cite{White92,White93}. The inherent efficiency of the DMRG algorithm lies in its reliance on the reduced density matrix to capture the essential quantum information, specifically, the entanglement between the system and its environment. By retaining only the reduced density-matrix eigenstates corresponding to the largest eigenvalues, the method effectively optimizes the basis and minimizes computational truncation error. The tensor network formalism is often preferable to traditional matrix-based packages primarily because it offers a unified, scalable, and computationally efficient representation of the many-body wave function, specifically the Matrix Product State (MPS). A key aspect of this methodology is the calculation of entanglement entropy, which is directly obtained from the singular value decomposition (SVD) between the tensor bonds,  an intrinsic step in the DMRG/MPS optimization procedure~\cite{Schollwock}.

The paper is organized as follows. In Sec.~\ref{model and method}, we introduce the frustrated spin-1/2 Heisenberg ladder model and outline the DMRG framework employed for its numerical analysis. Section~\ref{results} presents the ground-state phase diagram as a function of the bond-anisotropy parameter $\alpha$ and the transverse magnetic field, together with the corresponding magnetization, bipartite entanglement entropy, and spin–spin correlation functions. Particular emphasis is placed on the detailed characterization of the individual quantum phases, including the Haldane-like regime and the Tomonaga–Luttinger liquid phase, as well as on the analysis of their critical properties. Finally, Sec.~\ref{conclusion} summarizes the main results and discusses their broader physical implications.

\section{Model and method}\label{model and method}
\begin{figure}[b!]
{\includegraphics[width=.95\columnwidth,trim=0.5cm 4.6cm 9.5cm 20.5cm, clip]{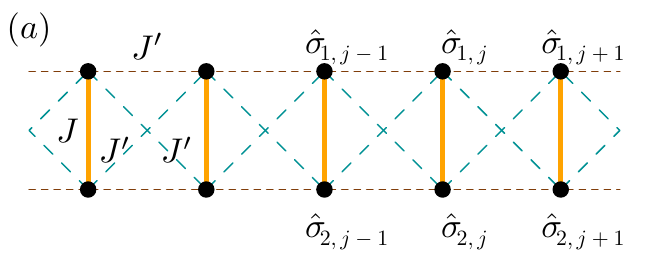}}\\
{\includegraphics[width=.95\columnwidth,trim=1.cm 4.6cm 8.2cm 21.5cm, clip]{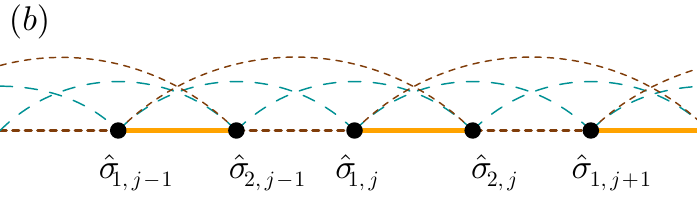}}
\caption{($a$) Schematic representation of the frustrated spin-$1/2$ two-leg ladder, consisting of $j=1,2,\dots,N/2$ vertical bonds with Heisenberg interaction $J$. The dashed horizontal and diagonal bonds denote inter-rung XXZ interactions $J^{\prime}$, comprising a longitudinal Ising component $J$ and a transverse exchange component $\alpha J$. 
($b$) One-dimensional enumeration of the two-leg ladder structure before transforming onto MPS, preserving the full interaction topology of the ladder.}
\label{fig0}
\end{figure}

We consider the spin-1/2 Heisenberg model on the two-leg ladder, as schematically shown in Fig.~\ref{fig0}. Each pair of vertically positioned spins $\hat{\boldsymbol \sigma}_{1, i}$ and $\hat{\boldsymbol \sigma}_{2, i}$ on the ladder forms a {\it rung}, where the two spins interact via the Heisenberg interactions $J$. Each rung is coupled to its nearest-neighbor rungs by four bonds---two horizontal and two diagonal---shown as dashed lines in Fig.~\ref{fig0}. These inter-rung bonds are anisotropic of the XXZ type: their longitudinal ($zz$) part keeps the fixed coupling $J$, whereas only their transverse ($xy$) part carries the tunable coupling $\alpha J$. Each inter-rung bond thus contributes $-J\hat{\sigma}^z_{~}\hat{\sigma}^z_{~}-\alpha J(\hat{\sigma}^x_{~}\hat{\sigma}^x_{~}+\hat{\sigma}^y_{~}\hat{\sigma}^y_{~})$, so that the pure Ising interaction $-J\hat{\sigma}^z_{~}\hat{\sigma}^z_{~}$ is recovered at $\alpha=0$ and the full isotropic Heisenberg interaction at $\alpha=1$. Hence, the parameter $\alpha$ gradually switches on the term $-\frac{\alpha J}{2}(\hat{\sigma}^+_{~}\hat{\sigma}^-_{~}+\hat{\sigma}^-_{~}\hat{\sigma}^+)$, where $\hat{\sigma}^{\pm}_{~}=\hat{\sigma}^{x}_{~}\pm i\hat{\sigma}^{y}_{~}$. We express the two-leg ladder Heisenberg model in the following form
\begin{align}
\label{Eq:hamiltonian}
\hat{H} = & -J \sum_{j = 1}^{\frac{N}{2}} \frac{1}{2} \left(
\hat{\sigma}^+_{1,j} \hat{\sigma}^-_{2,j}
+\hat{\sigma}^-_{1,j} \hat{\sigma}^+_{2,j}\right)
+\hat{\sigma}^z_{1,j} \hat{\sigma}^z_{2,j}
\nonumber\\
& -J \sum_{j = 1}^{\frac{N}{2}-1} \left(\hat{\sigma}^z_{1,j}+\hat{\sigma}^z_{2,j}\right)\left( \hat{\sigma}^z_{1,j+1}+\hat{\sigma}^z_{2,j+1}\right)
\nonumber\\
& - \frac{\alpha J}{2} \sum_{j = 1}^{\frac{N}{2}-1} \left( \hat{\sigma}^+_{1,j}+\hat{\sigma}^+_{2,j}\right)\left(\hat{\sigma}^-_{1,j+1}+\hat{\sigma}^-_{2,j+1} \right) \\ 
& - \frac{\alpha J}{2} \sum_{j = 1}^{\frac{N}{2}-1} \left( \hat{\sigma}^-_{1,j}+\hat{\sigma}^-_{2,j}\right)\left(\hat{\sigma}^+_{1,j+1}+\hat{\sigma}^+_{2,j+1} \right) \nonumber \\ 
& -h_x \sum_{j = 1}^{\frac{N}{2}} \left( \hat{\sigma}_{1,j}^{x} + \hat{\sigma}_{2,j}^{x}\right)
-h_z \left( \hat{\sigma}_{1,1}^{z} - \hat{\sigma}_{2,\frac{N}{2}}^{z}\right)
\nonumber \, ,
\end{align} 
where we consider the antiferromagnetic coupling constant $J<0$, a uniform magnetic field $h_x$ imposed along the $x$ direction, and a small magnetic field $h_z$ in the $z$ direction that is imposed on the first spin $\hat{\sigma}_{1,1}^{z}$ and the last spin $\hat{\sigma}_{2, N/2}^{z}$ only ($h_z$ is a small magnetic field that we use to break the symmetry). Later on, we consider higher values of $\alpha > 1$ to reach the XY model limit, where $\alpha\to\infty$.

\begin{figure}[!t]
{\includegraphics[width=1.\columnwidth,trim=0.cm 0.5cm 7cm 26cm, clip]{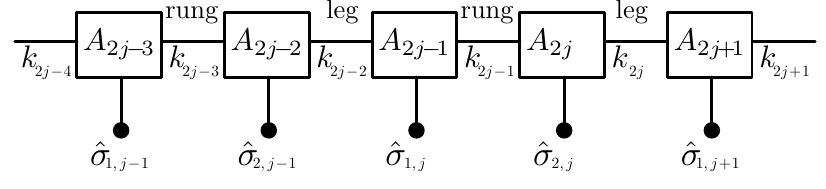}}
\caption{Graphical interpretation of MPS in Eq.~\eqref{mps}. Each rank-3 tensor $A$ is represented as a rectangle with three lines (indices): the vertical line ending with the filled circle is the physical 2-state variable $\sigma$ and the two horizontal lines are the $\chi$-state indices $k$ corresponding to the parameter $\chi$ known as the bond dimension. The horizontal indices are summed, and we name them alternately the {\it rung} and {\it leg} bonds. These labels denote the two inequivalent MPS bipartitions induced by the one-dimensional ordering of ladder sites shown in Fig.~\ref{fig0}($a$) and should not be confused with individual physical rung or leg bonds of the original ladder geometry. }
\label{fig0a}
\end{figure}

Having applied the DMRG method formulated in the MPS using the ITensor package~\cite{JuliaDMRG}, we solve the eigen-equation $\hat{H} |\psi_0\rangle = E_0 |\psi_0\rangle$ to find the ground state $|\psi_0\rangle$ and the corresponding lowest energy $E_0$ for the Hamiltonian in Eq.~\eqref{Eq:hamiltonian}. We calculate magnetization, von Neumann entanglement entropy, and the correlation functions from the MPS ground state for $N$ spins
\begin{equation}
    \vert\psi_0\rangle = \sum\limits_{\substack{\sigma_{1,1}^{~}\sigma_{1,2}^{~}\cdots\sigma_{1,N/2}^{~} \\ \sigma_{2,1}^{~}\sigma_{2,2}^{~}\cdots\sigma_{2,N/2}^{~}}}
    T^{\sigma_{1,1}^{~}\sigma_{1,2}^{~}\cdots\sigma_{1,N/2}^{~}}_{\,\sigma_{2,1}^{~}\sigma_{2,2}^{~}\cdots\sigma_{2,N/2}^{~}}
    \Big\vert {\substack{\sigma_{1,1}^{~}\sigma_{1,2}^{~}\cdots\sigma_{1,N/2}^{~} \\ \sigma_{2,1}^{~}\sigma_{2,2}^{~}\cdots\sigma_{2,N/2}^{~}}} \Big\rangle\, ,
\end{equation}
where the rank-$N$ tensor $T$ is written as the tensor product of $N$ rank-3 tensors $A$ of size $2\times\chi \times \chi$
\begin{align}
    \nonumber
    T^{\sigma_{1,1}^{~}\cdots\sigma_{1,N/2}^{~}}_{\,\sigma_{2,1}^{~}\cdots\sigma_{2,N/2}^{~}} & \approx \sum\limits_{k_0 \dots k^{~}_{N}}^\chi \prod\limits_{j=1}^{\frac{N}{2}} \left[A_{2j-1}\right]_{k_{2j-2}^{~}k_{2j-1}^{~}}^{\sigma^{~}_{1,j}} \left[ A_{2j} \right]_{k_{2j-1}^{~}k_{2j}^{~}}^{\sigma^{~}_{2,j}} \\
    & = \sum\limits_{k_0 \dots k^{~}_{N}}^\chi \prod\limits_{j=1}^{\frac{N}{2}} \prod\limits_{\ell=1}^{2}
    \left[A_{2j+\ell-2} \right]_{k_{2j+\ell-3}^{~}k_{2j+\ell-2}^{~}}^{\sigma^{~}_{\ell,j}}\,.
    \label{mps}
\end{align}
We introduced an integer variable $\chi$ known as the bond dimension $\chi \ll 2^{N/2}$. If $|\psi_0 \rangle$ is not a strongly correlated gapped state, it suffices to consider the bond dimension $1 \leq \chi \lesssim 10^3$. The vector basis has the standard direct product form
\begin{align}
    \nonumber
    \Big\vert {\substack{\sigma_{1,1}^{~}\sigma_{1,2}^{~}\cdots\sigma_{1,N/2}^{~} \\ \sigma_{2,1}^{~}\sigma_{2,2}^{~}\cdots\sigma_{2,N/2}^{~}}} \Big\rangle
     = 
    \bigotimes_{j=1}^{\frac{N}{2}} \Big\vert {\substack {\sigma_{1,j}^{~} \\ \sigma_{2,j}^{~}}} \Big\rangle 
    = \bigotimes_{j=1}^{\frac{N}{2}} \bigotimes_{\ell=1}^{2} |\sigma_{\ell,j}^{~}\rangle \, .
\end{align}
In Fig.~\ref{fig0a}, the two-state spin variables $\sigma_{\ell,j}^{~}$ denote the physical spins, whereas the $\chi$-state variables $k_{0}^{~},k_{1}^{~},\dots,k_{N}^{~}$ describe the bond interaction between the neighboring tensors $A$. We assume open boundary conditions throughout the work, such that the boundary bond indices are fixed to $k_0 = k_N = 1$, while only the internal bond indices carry the dimension $\chi$. The higher the bond dimension $\chi$, the more accurately the coefficients in the tensor $T$ approximate the ground state $|\psi_0\rangle$.
The numerical accuracy is controlled by setting the bond dimension $\chi$, which can take a nonzero integer value. This integer value refers to the maximum allowed reduced density-matrix states, i.e., the degrees of freedom of the bond variables $k_0^{~}, k_1^{~}, \dots, k_N^{~}$ in each tensor $A$. 

We calculate the spatial position-dependent magnetization
\begin{equation}
    \begin{split}
     \label{Eq:magn}
        \langle m_{\ell,j}^{\gamma} \rangle & = \langle \psi_0 | \hat{\sigma}_{\ell,j}^{\gamma} | \psi_0 \rangle
    \end{split}
\end{equation}
and the net magnetization per dimer 
\begin{equation}
    \begin{split}
    \label{Eq:magr}
        \langle m^{\gamma}_{~}\rangle & = \frac{2}{N} \sum\limits_{j=1}^{N/2} \left( \left\langle m_{1,j}^{\gamma} \right\rangle + \left\langle m_{2,j}^{\gamma} \right\rangle \right) \,  ,
    \end{split}
\end{equation}
where $\gamma=x,y,z$.
We can also evaluate von Neumann entanglement entropy by bipartitioning $|\psi_0\rangle$ between any nearest-neighboring spins ${\hat{\sigma}}^{~}_{\ell,j}$ and ${\hat{\sigma}}^{~}_{3-\ell,j+\ell-1}$ (where $\ell=1$ or $2$) reads
\begin{equation}
\label{Eq:entropy}
S_{\ell,j} = - \sum_{k_{2j+\ell-2}^{~}=1}^{\chi} s_{k_{2j+\ell-2}^{~}}^2 \ln s_{k_{2j+\ell-2}^{~}}^2\, ,
\end{equation}
where $s^{~}_{k_{2j+\ell-2}^{~}}$ are singular values after the SVD of $|\psi_0\rangle$.

We use the singular values, calculated at the ladder center ($j=N/4$ and $\ell=1$ or $2$), to estimate the efficiency and numerical accuracy of DMRG. To keep the reliable and accurate results, we consider a large bond dimension $100 < \chi < 2000$ and a sufficiently large number of sweeps (from 100 up to 200) to reach the full numerical convergence~\cite{Schollwock}, where we maintain the truncation error $ \varepsilon^{~}_{\ell}$ as small as possible, typically,
\begin{align}
   \varepsilon^{~}_{\ell} = 1-\sum_{k_{N/2+\ell-2}=1}^{\chi} s^{2}_{k_{N/2+\ell-2}} \lesssim 10^{-14}
\end{align}
for either $\ell=1$ or $2$ at the center of the ladder ($j=N/4$).

We thus distinguish bipartitioning of the ladder either between the two vertically oriented spins ${\hat{\sigma}}^{~}_{1,j}$ and ${\hat{\sigma}}^{~}_{2,j}$ if $\ell=1$, or between the diagonally adjacent spins ${\hat{\sigma}}^{~}_{2,j}$ and ${\hat{\sigma}}^{~}_{1,j+1}$ if $\ell=2$, as dictated by the MPS site ordering shown in Fig.~\ref{fig0}($b$). These correspond to the two inequivalent bipartitions of the MPS representation. The former defines the {\it rung} entanglement entropy $S_{\rm rung}$, whereas the latter defines the {\it leg} entanglement entropy $S_{\rm leg}$. In the thermodynamic limit ($N\to\infty$), the translational invariance of the entanglement entropy means that $S_{\ell=1,j} = S_{\ell=1,j+1}$ and $S_{\ell=2,j} = S_{\ell=2,j+1}$. However, we can only partially recover translational invariance near the ladder center, $ j = N/4$, unless $N$ is sufficiently large. We, therefore, focus on the calculation of rung entanglement entropy, $S_{\rm rung}$, (obtained between the spins ${\hat{\sigma}}^{~}_{1, N/4}$ and ${\hat{\sigma}}^{~}_{2, N/4}$) at the center of the ladder, and the nearest leg entanglement entropy, $S_{\rm leg}$ between ${\hat{\sigma}}^{~}_{2, N/4}$ and ${\hat{\sigma}}^{~}_{1, N/4+1}$
\begin{equation}
 \begin{split}
    \label{Eq:entropy2}
    S_{\rm rung} & \equiv S_{\ell=1,j=N/4}\,,\\
    S_{\rm leg} & \equiv S_{\ell=2,j=N/4} \,.
     \end{split}
\end{equation}
This classification helps us clearly determine partial and full separability, including locally entangled pairs, singlets, triplets, and multi-spin entangled states, using magnetization obtained from Eqs.~\eqref{Eq:magn}. 

To provide a more detailed specification of the model, we also evaluate the nearest-neighbor correlation functions. Analogously to the entanglement entropy, we define the nearest-neighbor correlation function
\begin{equation}
        C_{\ell,j}^{\gamma} = \langle \psi_0 | \hat{\sigma}_{\ell,j\ }^\gamma \hat{\sigma}_{3-\ell,j+\ell-1}^\gamma |\psi_0 \rangle\, ,
\end{equation}
where $\gamma=x,y,z$ from which we calculate the correlation function
on the rung $C_{\rm rung}$ (i.e. $\ell=1$) and between two spins on the neighboring rungs $C_{\rm leg}$ (i.e., $\ell=2$)
\begin{equation}
    \begin{split}
    \label{Eq:corr}
        C_{\rm rung}^{~} & = \sum\limits_{\gamma=x,y,z} C_{1,N/4}^{\gamma} = \sum\limits_{\gamma=x,y,z} \langle \psi_0 | \hat{\sigma}_{1,N/4\ }^\gamma \hat{\sigma}_{2,N/4}^\gamma |\psi_0 \rangle \, ,\\
        C_{\rm leg}^{~} & =\sum\limits_{\gamma=x,y,z} C_{2,N/4}^{\gamma} = \sum\limits_{\gamma=x,y,z} \langle \psi_0 | \hat{\sigma}_{2,N/4\ }^\gamma \hat{\sigma}_{1,N/4+1}^\gamma|\psi_0 \rangle\, .
    \end{split}
\end{equation}

To characterize the critical behavior and universality class of the quantum phase transitions, we extract the central charge $c$ from the finite-size scaling of the rung entanglement entropy. At criticality, $S_{\rm rung}(N)$ follows the conformal field theory form $S_{\rm rung}(N) = S_0 + \tfrac{c}{6}\ln N$ for open boundary conditions, evaluated at the system center~\cite{Calabrese}. We determine $c$ and $S_0$ as the free fitting parameters for the numerical data of $S_{\rm rung}$ using this scaling relation. The central charge reflects the number of effective low-energy degrees of freedom and distinguishes critical from noncritical ground states.

\section{Results}\label{results} 
\subsection{Ground-state phase diagram}
\label{MagOrd}
\begin{figure}[!t]
\includegraphics[width=1.0\columnwidth,clip]{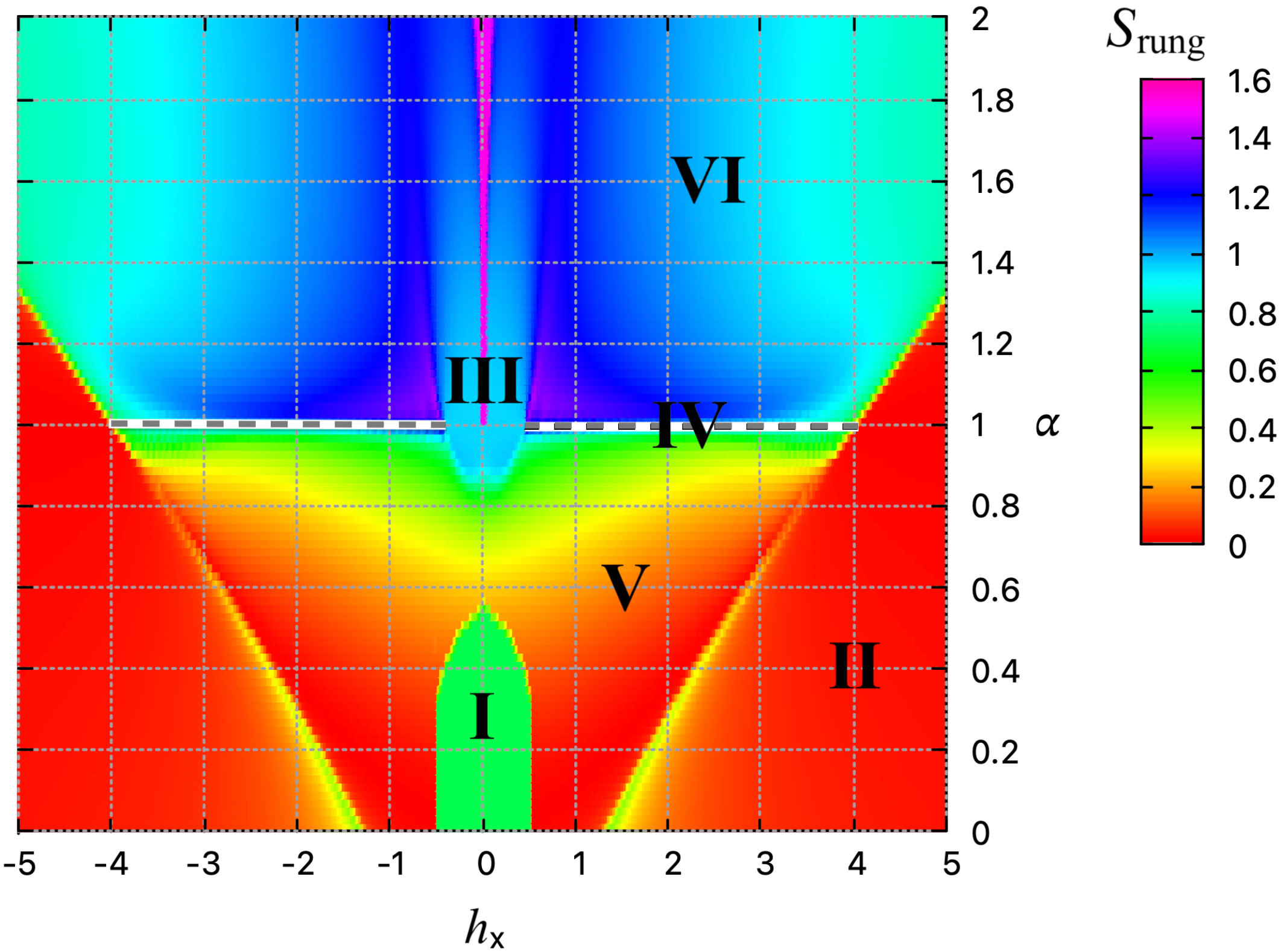}
\caption{Schematic phase diagram of the model defined in Eq.~\eqref{Eq:hamiltonian} where we plot $S_{\rm rung}$ versus $h_x$ and $\alpha$. The notation is as follows: I—rung-singlet phase; II—ordered ferromagnetic phase; III—Haldane-like phase; IV—Luttinger liquid (gray dashed thick line); V—canted Ising phase; and VI—XY-polarized phase. The DMRG results are obtained for $N=100$ using $200$ sweeps, with a truncation error $\varepsilon_{1}^{~} \lesssim 10^{-14}$.  }
\label{fig:fd}
\end{figure}
Fig.~\ref{fig:fd} shows the phase diagram over the full parameter space spanned by $h_x$ and $\alpha$, identifying six distinct ground-state phases $\vert \psi_0\rangle$. Their classification is based on the behavior of the onsite $z$-magnetization, $\langle m^{z}_{\ell,j}\rangle$ in Eq.~\eqref{Eq:magn}, the net $x$-magnetization per dimer, $\langle m^x_{~}\rangle$ in Eq.~\eqref{Eq:magr}, and entanglement entropy on the rung $S_{\rm rung}$ and on the leg $S_{\rm leg}$ in Eq.~\eqref{Eq:entropy2}. This analysis is further supported by the corresponding nearest-neighbor spin–spin correlation functions $C_{\rm rung}$ and $C_{\rm leg}$ in Eq.~\eqref{Eq:corr}. We also perform the finite-size scaling of entanglement entropy, from which the central charge $c$ is extracted to distinguish between critical and noncritical phases. The defining characteristics of each phase are summarized in Tab.~\ref{tab:tab1}, while their regions of stability are indicated in Fig.~\ref{fig:fd}. Since all analyzed physical quantities are symmetric with respect to $\pm h_x$, we restrict the discussion to $h_x\geq 0$. We set $J=-1$ in the numerical calculations that corresponds to the antiferromagnetic coupling.

\begin{figure*}[t!]
{\includegraphics[width=1\columnwidth,clip]{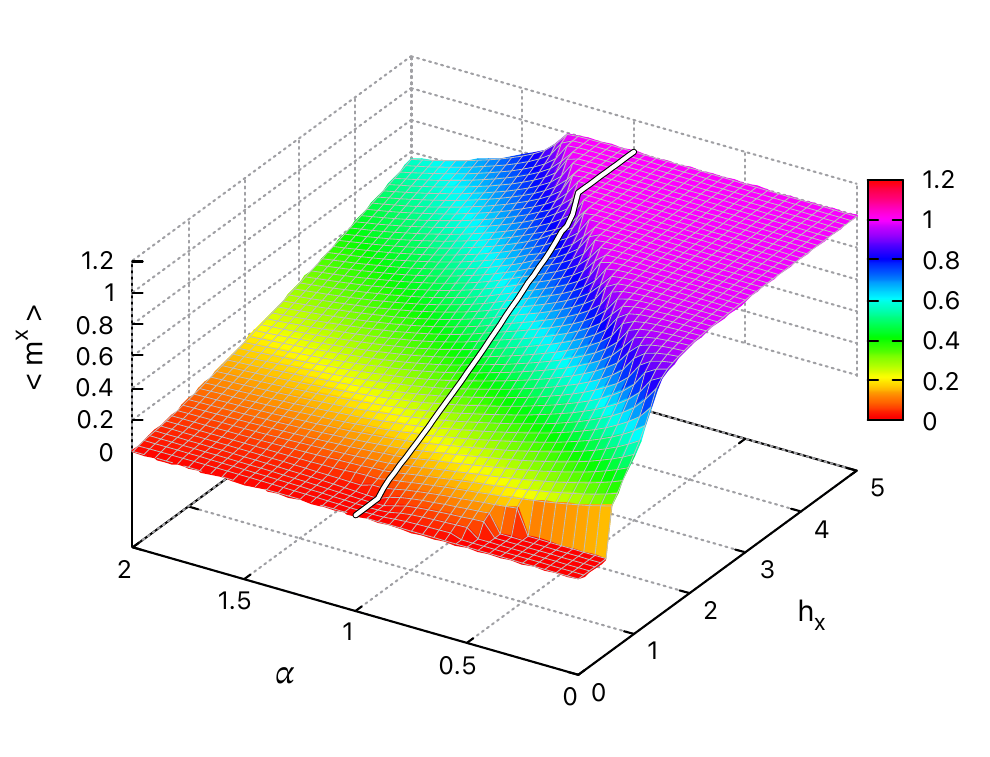}}
{\includegraphics[width=1\columnwidth,clip]{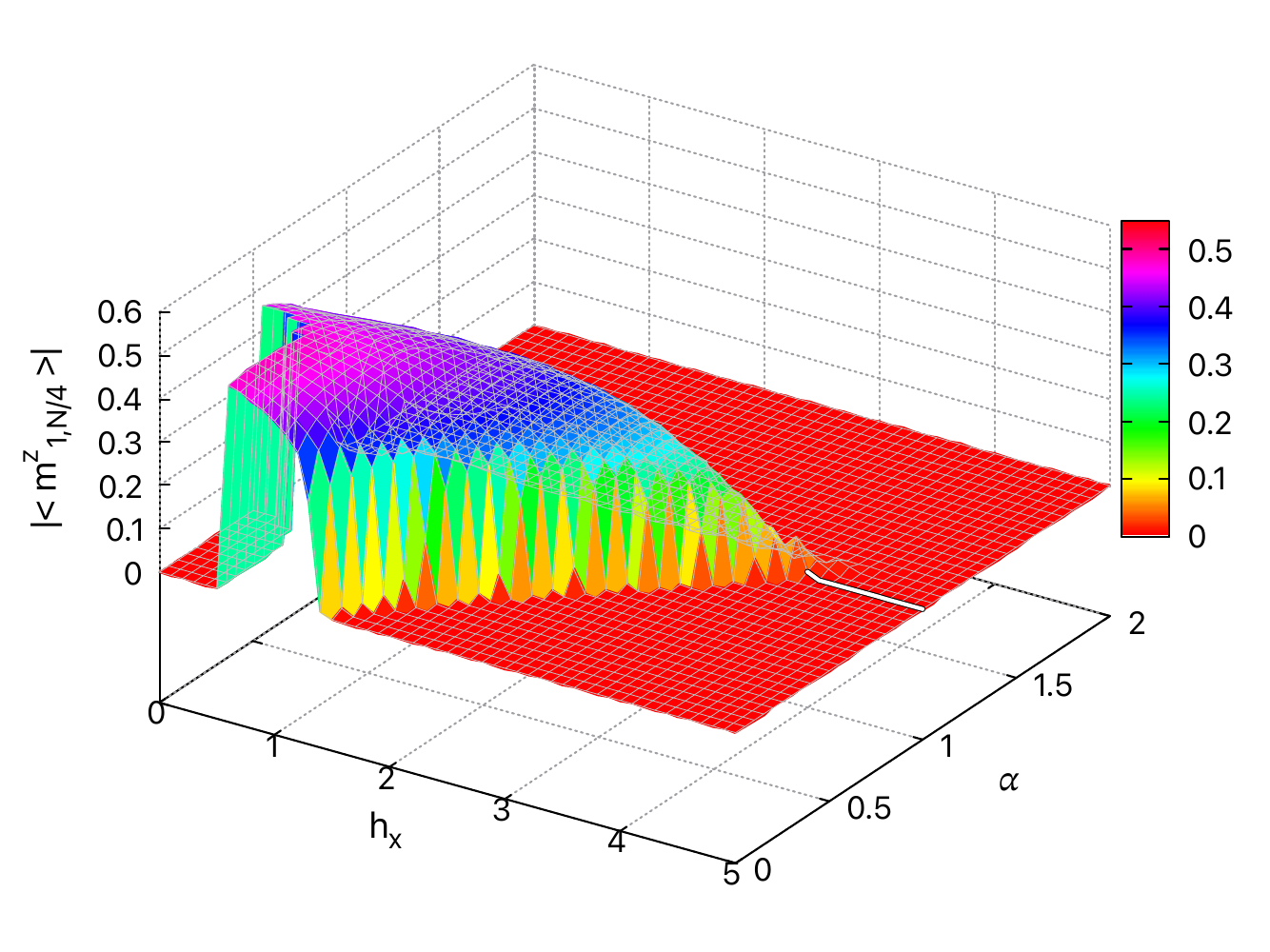}}
\caption{3D plot of the magnetization as a function of $h_x$ and $\alpha$, for a system size of $N=100$. The thick white curve corresponds to the isotropic case $\alpha=1$. Left panel: net magnetization per dimer $\langle m^x\rangle$ defined in Eq.~\eqref{Eq:magr}. Right panel: magnitude of the onsite magnetization $|\langle m^z_{\ell=1,j=N/4}\rangle|$ at the central site defined in Eq.~\eqref{Eq:magn}.}
\label{fig:mag}
\end{figure*}

\renewcommand{\arraystretch}{1.5}
\begin{table}[b!]
    \begin{center}
        \begin{tabular}{|c | l | c | c | c | c | c | c | c | c |} 
        \hline
        \# & \ Phase type & $\langle m^x_{~}\rangle $ & $\langle m^z_{\ell,j}\rangle $& $S_{\rm rung}$ & $S_{\rm leg}$ & $C_{\rm rung}$ & $C_{\rm leg}$ & $c$ \\ [0.5ex]   \hline\hline
        I & Rung singlet &  $\phantom{\approx}\,\,0$ & $\phantom{\neq}\,\, 0$ & $\ln{2}$ & $\phantom{\approx}\,\,0$ & $-\tfrac{3}{4}$ & $\phantom{\lessapprox} 0$&0\\ 
        \hline
        II & Ordered ferro &  $\lessapprox 1$ & $\phantom{\neq}\,\, 0$ & $\approx 0$& $\approx 0$ & $\phantom{-}\tfrac{1}{4}$ & $\lessapprox  \tfrac{1}{4}$&0\\ 
        \hline
        III & Haldane-like  & $\approx 0$ & $\phantom{\neq}\,\, 0$ & $> \ln{2}$ & $> \ln{2}$ & $\phantom{-}\tfrac{1}{4}$& $ \approx a$& 0\\
        \hline
        IV & Luttinger liq. & $\neq0$ & $=0$ & $> \ln{2}$ & $> \ln{2}$ & $\phantom{-}\tfrac{1}{4}$&$[a,\tfrac{1}{4}]$ & 1\\
        \hline 
        V & Canted Ising & $\neq0$ & $\neq0$ & $\lesssim\ln{2}$ & $\lesssim\ln{2}$  & $\phantom{-}\tfrac{1}{4}$ &$[a,\tfrac{1}{4}]$& 0 \\ 
        \hline
        VI & XY-polarized & $\neq0$ & $=0$ & $>\ln{2}$ & $>\ln{2}$  & $\phantom{-}\tfrac{1}{4}$ &$[a,\tfrac{1}{4}]$& 0 \\ 
        \hline
        \end{tabular}
    \end{center}
    \caption{The list of the six ground-state phases characterized by the magnetizations $\langle m^x_{~}\rangle$, $\langle m^z_{\ell,j}\rangle$, entanglement entropy $S_{\rm rung}$, $S_{\rm leg}$, the spin-spin correlation functions $C_{\rm rung}$, $C_{\rm leg}$, and the central charge $c$. The constant $a=\min_{[\alpha,h_x]}(C_{\rm leg})=-0.350371$ was determined at $\alpha=1$ and $h_x=0$ in the thermodynamic limit ($N\to\infty$).
    }
    \label{tab:tab1}
\end{table}

\subsection{Magnetic properties}
\label{MagProp}

 We first calculate the net magnetization  $\langle m^x \rangle$ shown in the left panel of Fig.~\ref{fig:mag} as a function of the magnetic field $h_x$ and the anisotropy parameter $\alpha$. The resulting surface reveals three qualitatively distinct regimes governed by the interplay between exchange anisotropy and transverse-field-induced quantum fluctuations.
 
In the strong-field limit, the system approaches an ordered ferromagnetic state with $\langle m^x \rangle \to 1^{-}$, corresponding to nearly parallel alignment of all spins along the field direction. For the isotropic case $\alpha = 1$ (highlighted by the thick white curve), this regime is preceded by a well-defined magnetization plateau. Upon introducing exchange anisotropy, $\alpha \neq 1$, this plateau progressively transforms into a quasi-plateau, with saturation reached only asymptotically as $h_x \to \infty$. Microscopically, this evolution reflects the competition between suppressed transverse fluctuations in the Ising-like regime for $\alpha < 1$ and enhanced inter-rung XY processes for $\alpha > 1$, which favor a gradual and correlated polarization. In the thermodynamic limit, these quasi-plateaus correspond to continuously evolving, partially polarized states rather than truly gapped plateaus.

At intermediate fields, the magnetization varies continuously with the applied field over a broad parameter range, indicating the absence of stable plateaus and the presence of smoothly evolving spin configurations. The extent of this regime is strongly controlled by $\alpha$: for $\alpha \gtrsim 0.5$ and $\alpha \neq 1$, the continuous response sets in already at very small fields ($h_x \to 0$), whereas in the limiting cases $\alpha \to 0$ and $\alpha \to 1$, a finite threshold field of order $h_x \sim 0.5$ is required. 

The low-field regime further differentiates the 3D magnetization diagram. The zero-magnetization plateau is observed if $0 \leq \alpha \lesssim 0.5$, corresponding to a gapped phase. At its boundary, the magnetization exhibits a discontinuous jump to a finite value, indicating a first-order phase transition. By contrast, at the isotropic point $\alpha = 1$, a second zero-magnetization plateau vanishes continuously with increasing field, signaling a transition into a gapless canted phase belonging to the commensurate-incommensurate universality class~\cite{Fouet}. In the vicinity of $\alpha = 1$, the magnetization remains strongly suppressed, $\langle m^x\rangle \sim 10^{-4}$--$10^{-6}$, and increases gradually with $h_x$ and $\alpha$, consistent with proximity to this continuous transition.

Notably, the net longitudinal component $\langle m^z \rangle$ vanishes identically throughout the entire parameter space. Meanwhile, the onsite $z$-magnetization, $\langle m^z_{\ell,j} \rangle$ (right panel of Fig.~\ref{fig:mag}), remains finite for $\alpha < 1$ up to the intermediate-field regime and exhibits a collinear arrangement along the $z$-direction, characterized by ferromagnetic alignment within each rung and an alternating orientation of neighboring rungs. Its maximum value of $\langle m^z_{1,N/4} \rangle = \tfrac{1}{2}$ is gradually suppressed with increasing $h_x$ and $\alpha$, as illustrated in the right panel of Fig.~\ref{fig:mag}. Although finite-size calculations yield a small nonzero value of $\langle m^z_{\ell,j} \rangle$ within the Luttinger-liquid phase, a finite-size scaling analysis indicates that this quantity vanishes in the thermodynamic limit. Consequently, the regime $\alpha < 1$ excludes the critical point $\alpha = 1$. 

\subsection{Entanglement entropy}
\label{EntEnt}

\begin{figure}[!t]{\centering\includegraphics[width=1\columnwidth,clip]{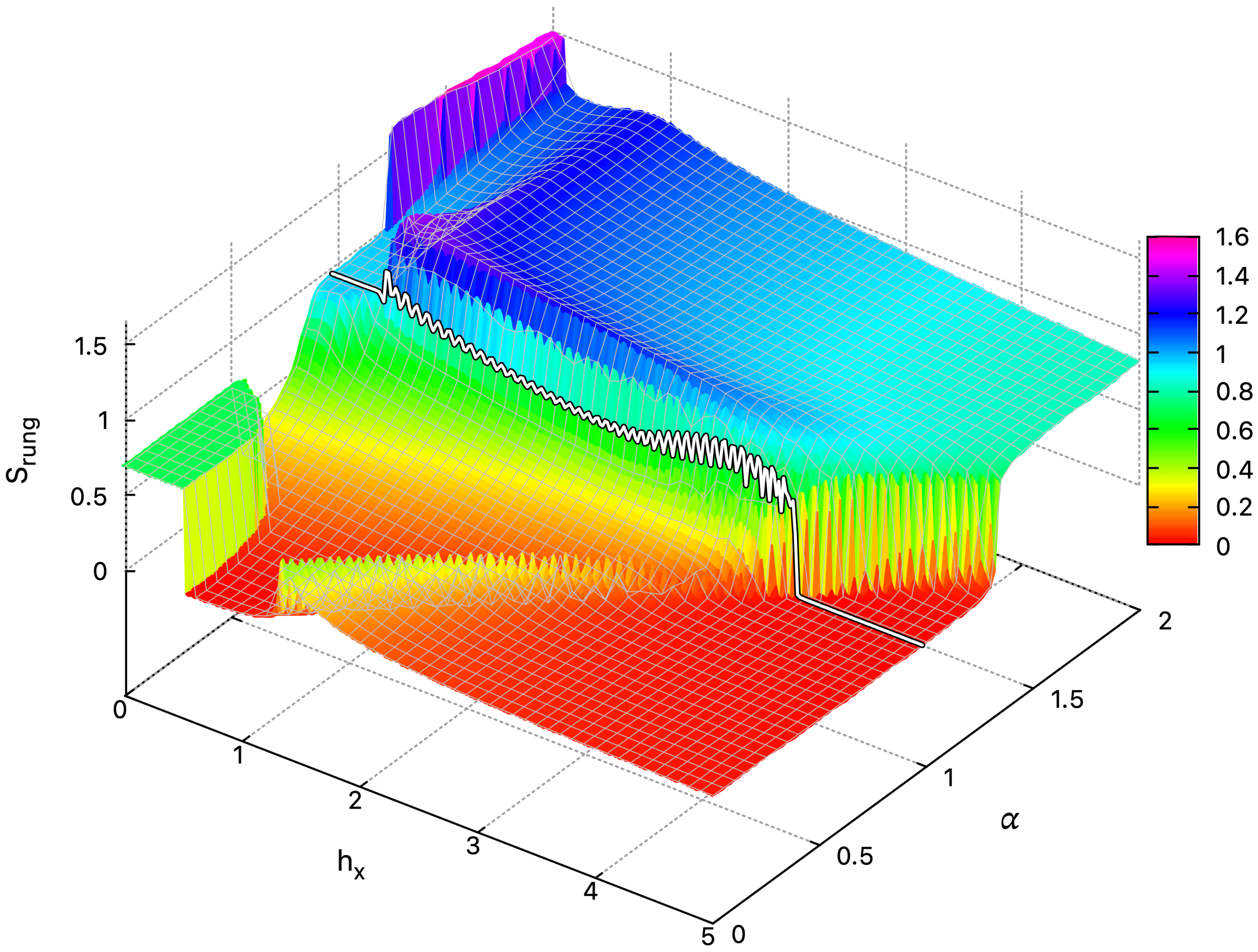}}
  \caption{3D plot of the von Neumann entanglement entropy $S_{\rm rung}$ defined in Eq.~\eqref{Eq:entropy2} as a function of $h_x$ and $\alpha$, for $N=100$. The thick white curve corresponds to the isotropic case $\alpha=1$. }
\label{fig:ent}
\end{figure}

Fig.~\ref{fig:ent} illustrates the von Neumann rung entanglement entropy, $S_{\rm rung}$, as a function of $h_x$ and $\alpha$, calculated according to Eq.~\eqref{Eq:entropy2}. The thick white curve highlights the fully isotropic case $\alpha=1$. We find that $S_{\rm rung}$ and $S_{\rm leg}$  exhibit nearly identical behavior over most of the parameter space, except for the region $0\leq h_x\lesssim 0.5$ and $0\leq \alpha\lesssim 0.5$, reflecting a redistribution of correlations across the ladder.
\begin{figure*}[!bth]
{\centering\includegraphics[width=1\columnwidth,clip]{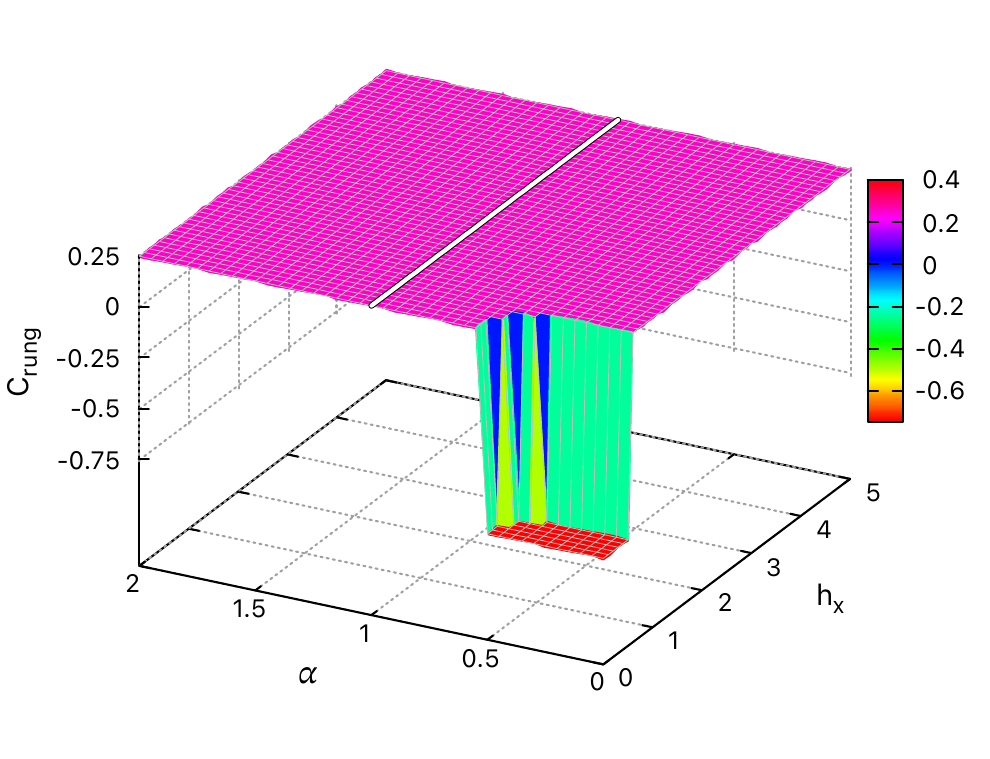}}
{\centering\includegraphics[width=1\columnwidth,clip]{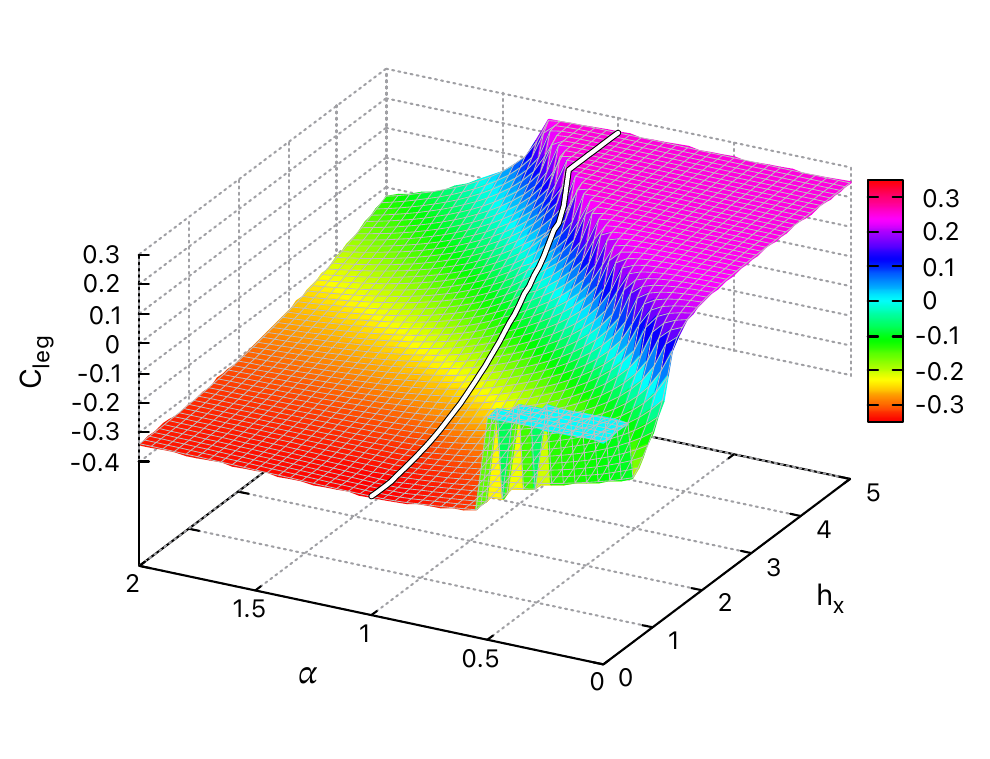}}
  \caption{3D plot of the nearest-neighbor spin-spin correlation functions, defined by  Eq.~\eqref{Eq:corr}, as a function of $\alpha$ and $h_x$ calculated for $N=100$. The thick white curve indicates the correlation function at $\alpha=1$. Left panel: rung correlation function $C_{\rm rung}$. Right panel: leg correlation function $C_{\rm leg}$.}
\label{fig:corr}
\end{figure*}
In this regime, the spins form maximally entangled rung dimers with $S_{\rm rung}=\ln 2$, visible as a pronounced plateau in the lower-left corner of Fig.~\ref{fig:ent}. Meanwhile, $S_{\rm leg} = 0$, indicating the absence of inter-rung entanglement and a separable structure composed of rung dimers. Both entropies remain constant throughout this region, demonstrating the robustness of the dimerized structure against weak transverse fields ($0\leq h_x\lesssim0.5$).

A nearly linear boundary (from $h_x\approx 1.3$ at $\alpha=0$ to $h_x=5$ at $\alpha=4/3$) separates two additional regimes. Beyond this boundary, both $S_{\rm rung}$ and $S_{\rm leg}$ decrease continuously toward zero, reflecting the suppression of quantum correlations as spins progressively align with the field $h_x$. On the other side of the boundary, a clear distinction emerges between the regimes $\alpha < 1$ and $\alpha > 1$, with the isotropic case $\alpha = 1$ marking the crossover between them. For $\alpha < 1$, the entanglement entropy remains finite and typically below $\ln 2$, with a pronounced dependence on $\alpha$: it decreases as $\alpha \to 0$ and increases toward $\ln 2$ as $\alpha \to 1$, consistent with the evolution from weakly correlated to maximally entangled rung dimers. In contrast, for $\alpha > 1$ the entanglement entropy exceeds $\ln 2$ and exhibits a strong dependence on the transverse field $h_x$, indicating enhanced quantum correlations beyond the simple dimer picture. In this regime, $S_{\rm rung}$ increases with anisotropy and approaches values consistent with an effective enlargement of the local rung Hilbert space, tending toward $\ln 4$ in the strong-anisotropy limit ($\alpha \to \infty$). 

Finally, in the vicinity of $\alpha \to 1$ and $h_x \to 0$, a small region with $S_{\rm rung}>\ln 2$ and $S_{\rm leg}>\ln 2$ emerges despite a strongly suppressed magnetization $\langle m^x\rangle$. While the magnetic response remains nearly indistinguishable from that of the separable dimer regime ($0\leq h_x \lesssim 0.5$ and $0\leq \alpha \lesssim 0.5$), the substantially enhanced entanglement entropy indicates a fundamentally different pattern of quantum correlations. 

\subsection{Spin-spin correlation function}

Fig.~\ref{fig:corr} shows the behavior of the nearest-neighbor spin–spin correlation functions $C_{\rm rung}$ (left panel) and $C_{\rm leg}$ (right panel), calculated according to Eq.~\eqref{Eq:corr} under simultaneous variation of the model parameters $h_x$ and $\alpha$. The thick white curves in both panels highlight the isotropic case $\alpha = 1$.

The rung correlation $C_{\rm rung}$ assumes only two distinct values. Within the region of maximally entangled rung dimers ($0 \leq h_x \lesssim 0.5$ and $0 \leq \alpha \lesssim 0.5$), $C_{\rm rung} = -\frac{3}{4}$, whereas it takes the value $C_{\rm rung} = +\tfrac{1}{4}$ across the remainder of the parameter space. The abrupt transition between these values reflects the first-order nature of the corresponding phase boundary.

By contrast, $C_{\rm leg}$ exhibits a more intricate structure. It forms a zero-correlation plateau that perfectly coincides with the uncorrelated rung-dimer region, and a quasi-plateau at $C_{\rm leg} \approx +\tfrac{1}{4}$ within the ordered ferromagnetic phase ($\langle m^x \rangle \to 1^{-}$). Between these limits, $C_{\rm leg}$ varies continuously, evolving from $C_{\rm leg} = a\approx -0.350371$ for $h_x \to 0^{+}$ up to $C_{\rm leg} = +\tfrac{1}{4}$ near the dominant nearly linear phase boundary. Notably, the onset of the continuous regime from the zero-correlation plateau is discontinuous, further signaling a first-order transition, whereas the crossover from the quasi-plateau to the continuous regime remains smooth.

\subsection{Ground-state phase diagram}
\renewcommand\thesubsubsection{\Roman{subsubsection}}
\setcounter{subsubsection}{0}

Having characterized the global magnetic response, entanglement, and spin–spin correlations, we now examine each distinct ground state in the phase diagram (Fig.~\ref{fig:fd}) in detail. This phase-resolved analysis highlights the microscopic spin arrangements, correlation patterns, and entanglement structure that define the nature of each regime.

\subsubsection{Rung-singlet phase}
The rung-singlet phase, denoted as phase I in Fig.~\ref{fig:fd}, occupies the region $|h_x|\lesssim0.5$ and $0\leq \alpha \lesssim 0.5$. The onsite magnetization vanishes across the ladder, $\langle m_{\ell,j}^x \rangle = \langle m_{\ell,j}^z \rangle = 0$ for all $\ell=1,2$ and $1\leq j\leq N/2$, signaling the absence of spontaneous magnetic order and the lack of a preferred spin orientation. The rung entanglement entropy is constant at $S_{\rm rung}=\ln 2$, while the leg entanglement entropy vanishes, $S_{\rm leg}=0$, indicating that the rungs remain separable. 
The rung spin–spin correlation in the bulk attains the value $C_{\rm rung}=-\tfrac{3}{4}$, confirming the formation of singlets along the rungs. All other spin–spin correlations vanish, $C_{\ell=2,j}^{\gamma=x,y,z}=0$, demonstrating that the singlets are mutually uncorrelated and the ground state is a direct product of independent maximally entangled rung dimers. The strictly local nature of these correlations implies a finite spin gap separating the singlet ground state from the triplet excitations, as expected for a rung-dimerized ladder~\cite{Dagotto92, Chitra}. Overall, this phase represents a short-range entangled, gapped, quantum-disordered state with correlations confined to individual rungs.

\subsubsection{Ordered ferromagnetic phase}

The ordered ferromagnetic phase, denoted as phase II in Fig.~\ref{fig:fd}, occupies the triangular regions at the lower corners of the phase diagram, delimited by anisotropy-dependent transition lines separating it from neighboring phases. The phase boundary exhibits a distinct dependence on the anisotropy: for $\alpha > 1$ it follows a linear form $h_c^{(\mathrm{VI-II})} = 3\alpha + 1$, whereas for $\alpha < 1$ it becomes nonlinear, reflecting the enhanced competition between exchange anisotropy and the transverse field. Within this phase, the net magnetization $\langle m^x \rangle$ approaches its maximal value, $\langle m^x \rangle \to 1^{-}$, indicating a nearly fully polarized state with uniform alignment of all onsite magnetic moments along the applied transverse field. Simultaneously, the longitudinal magnetization vanishes at every site, $\langle m_{\ell,j}^z \rangle = 0$, reflecting the suppression of residual Ising-like correlations and the dominance of field-induced ferromagnetic order along the $x$-direction. This phase thus corresponds to a fully (field-)polarized product state. At the isotropic point $\alpha = 1$, the system becomes exactly fully polarized already at finite $h_x$, with $\langle m^x \rangle = 1$, which is an exact product-state  of $|\psi_0 \rangle$.

The entanglement entropies, $S_{\rm rung}$ and $S_{\rm leg}$, vanish in the limit $h_x \to \infty$, consistent with an asymptotically fully polarized product state. For $\alpha \neq 1$, finite entropies persist at intermediate fields, reflecting residual quantum correlations induced by competing noncommuting exchange interactions, which prevent exact factorization at finite field strength. In this regime, the ground state remains weakly entangled, with deviations from perfect polarization becoming more pronounced at intermediate fields. In contrast, at the isotropic point $\alpha = 1$, both entanglement entropies vanish throughout the ordered ferromagnetic regime, consistent with an exactly separable product-state ground state.

Spin–spin correlations are predominantly aligned along the $x$-direction for both rung and leg bonds, while correlations in orthogonal components are strongly suppressed. Both rung ($\ell=1$) and leg ($\ell=2$) correlations along $x$-direction reach $C_{\ell=1,j}^{\gamma=x} \simeq C_{\ell=2,j}^{\gamma=x} \simeq +\tfrac{1}{4}$, consistent with fully polarized $\vert\uparrow\uparrow\rangle$ triplet configurations. The uniform positive correlations confirm the establishment of long-range ferromagnetic order throughout the ladder, which becomes asymptotically classical in the large-field limit.

\subsubsection{Haldane-like phase}

\begin{figure}[bt!]
{\includegraphics[width=1\columnwidth,clip]{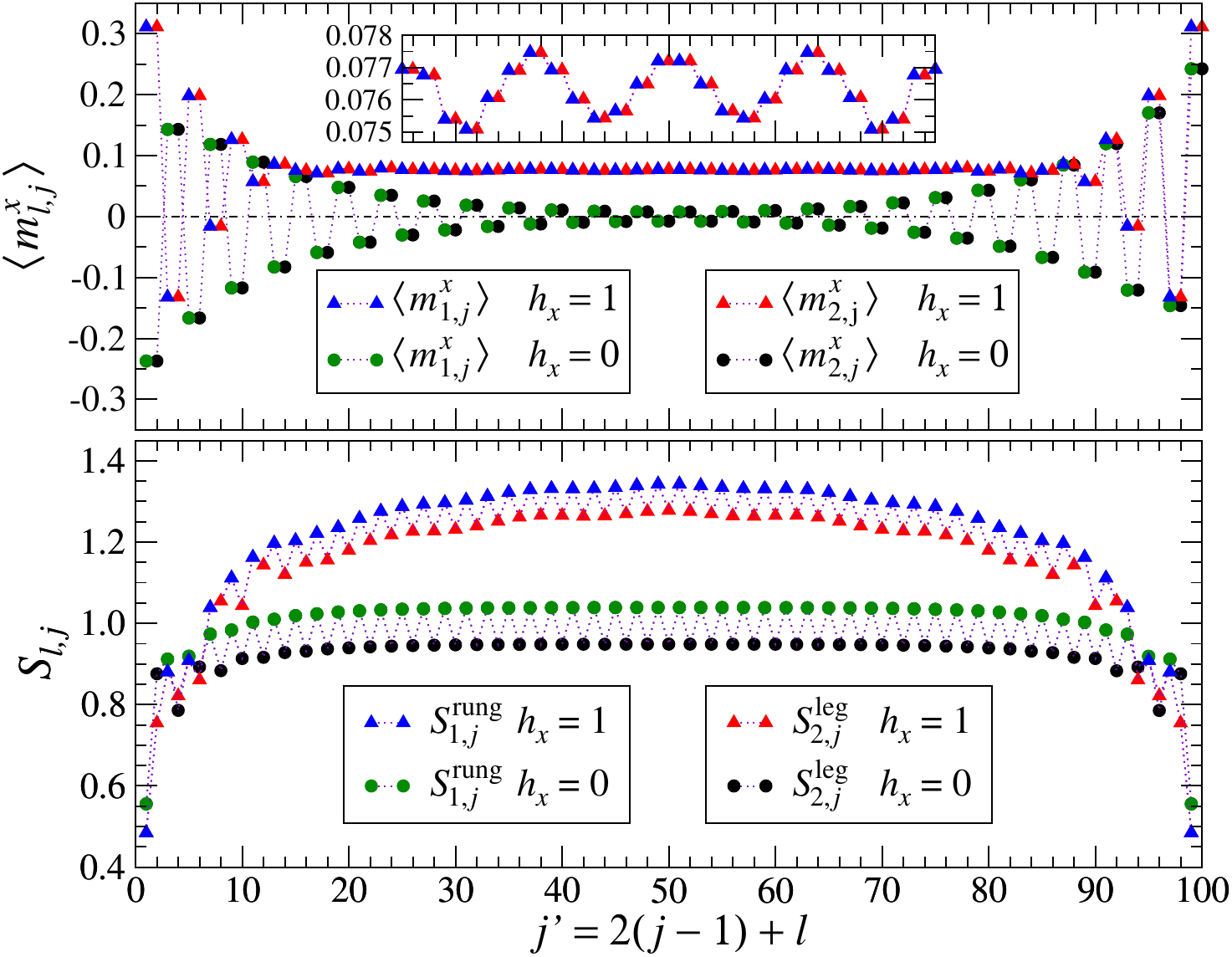}}
  \caption{Position dependence of the onsite magnetization $\langle m^x_{\ell,j} \rangle$ and the entanglement entropy $S_{\ell,j}$ on the rung ($\ell=1$) and leg ($\ell=2$) in the Haldane-like phase ($h_x=0$) and the Luttinger-liquid phase ($h_x=1$) for $\alpha=1$ and $N=100$, under open boundary conditions. Here $j=1,\dots,N/2$ for each leg.}
\label{fig:edge}
\end{figure}
The region denoted as phase III in Fig.~\ref{fig:fd} is more appropriately described as a Haldane-like regime, occurring in the vicinity of the isotropic point and weak transverse field ($|h_x|\lesssim0.44$, $\alpha \approx 1$). At the fully isotropic point $\alpha=1$, both net magnetizations, $\langle m^{x} \rangle$ and $\langle m^{z} \rangle$, exhibit well-defined zero plateaus. Upon introducing exchange anisotropy ($\alpha\neq1$), a small but finite transverse magnetization develops under the applied field, indicating a gradual departure from the isotropic limit. The onsite magnetization profiles exhibit a nontrivial spatial structure throughout this parameter region, characterized by alternating rung correlations and an enhancement of local moments toward the system boundaries. This behavior is consistent with effective spin-$1/2$ edge excitations in finite open ladders arising from an effective spin-1 description of the low-energy sector~\cite{Qin, Pasnoori}. The characteristic edge modes in the $x$ direction for a finite ladder of $N=100$ sites are illustrated in the upper panel of Fig.~\ref{fig:edge} for $\alpha=1$ and $h_x=0$.
\begin{figure}[!t]
{\includegraphics[width=\columnwidth]{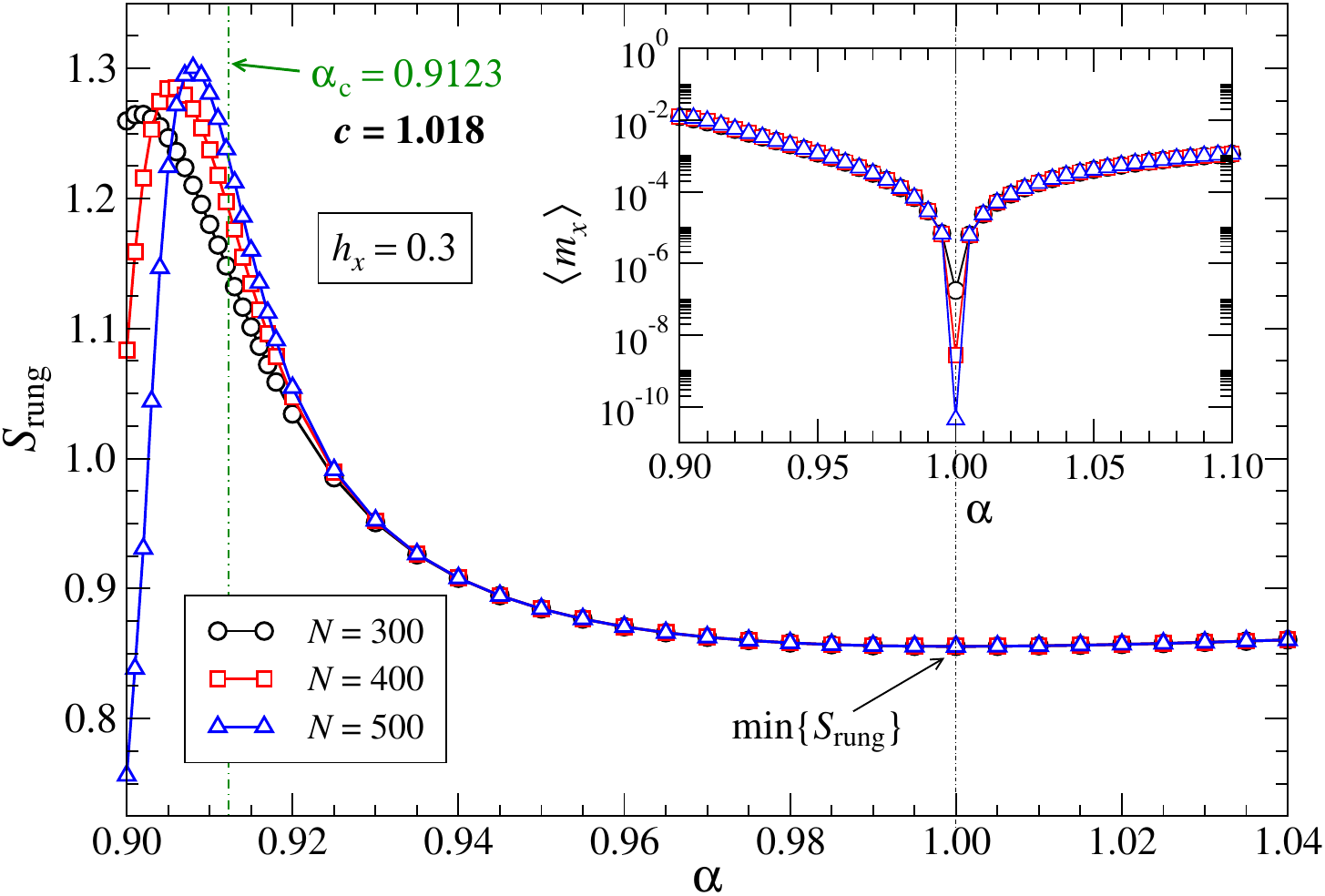}}
  \caption{Finite-size scaling of the rung entanglement entropy $S_{\rm rung}$ as a function of the exchange anisotropy $\alpha$ for $h_x=0.3$ and system sizes $N=300$, $400$, and $500$. Extrapolation of the entropy maxima to the thermodynamic limit yields the critical anisotropy $\alpha_c^{\rm (III-V)}=0.9123$ separating the canted Ising and Haldane-like phases. The corresponding finite-size scaling $S_{\rm rung}(\alpha_c)=S_0+\tfrac{c}{6}\ln N$ gives the central charge $c=1.018$ and $S_0=0.1748$. Inset: finite-size scaling of the net magnetization $\langle m^x\rangle$ for the same parameters, showing a pronounced minimum at $\alpha=1$.  }
\label{fig:Haldane_scale}
\end{figure}

\begin{figure}[t!]
{\centering\includegraphics[width=\columnwidth]{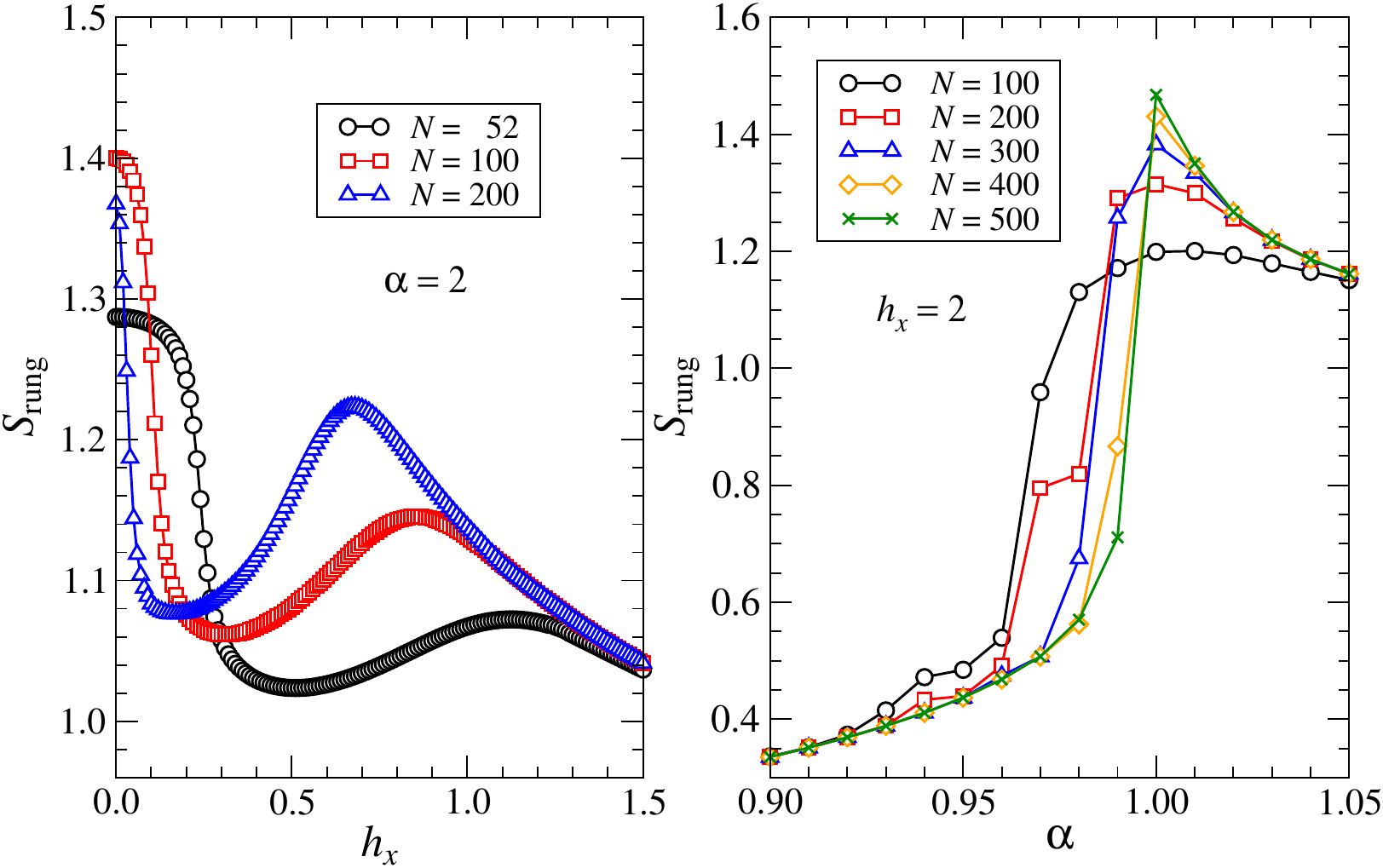}}
  \caption{  Left: Size dependence of $S_{\rm rung}$ as a function of the magnetic field $h_x$ for $\alpha = 2$ and $N = 52, 100,$ and $200$ (we demand divisibility of $N$ by 4). Right: Size dependence of $S_{\rm rung}$ near $\alpha = 1$ at $h_x = 2.0$ for $N = 100$-$500$, showing an increase with system size and a gradual suppression of the enhanced-entropy region, consistent with a Tomonaga–Luttinger liquid behavior in the isotropic limit. } 
\label{scale_S_hx2c0}
\end{figure}

Both $S_{\rm rung}$ and $S_{\rm leg}$ exceed $\ln 2$, reflecting a stronger entanglement beyond a simple rung-dimer picture in phase I. The entanglement entropies display a nearly uniform bulk profile with boundary-induced suppression near the edges (lower panel of Fig.~\ref{fig:edge}), consistent with the area-law behavior expected for gapped one-dimensional systems with open boundaries~\cite{Schollwock}. Additional insight is obtained from the finite-size scaling of the rung entanglement entropy $S_{\rm rung}$. As shown in Fig.~\ref{fig:Haldane_scale}, the enhanced-entanglement region remains stable over a finite interval around the isotropic point, with the results for $N=300$, $400$, and $500$ collapsing onto a common curve for $\alpha \gtrsim 0.93$. Simultaneously, the net transverse magnetization exhibits a pronounced minimum at $\alpha=1$ (inset of Fig.~\ref{fig:Haldane_scale}), where $\langle m^x\rangle$ decreases toward zero with increasing system size. These observations indicate that the Haldane-like characteristics are not restricted to the single isotropic point but persist within a narrow neighborhood of $\alpha=1$. The scaling behavior also shows a strong sensitivity to exchange anisotropy, with  a transition to the canted Ising phase occurring near $\alpha_c \approx 0.9123$. Finite-size scaling at this extrapolated critical point yields a central charge $c=1.018$ from $S_{\rm rung}(\alpha_c)=S_0+\tfrac{c}{6}\ln N$. This result is consistent with $c\simeq 1$ critical behavior, suggesting a Gaussian (Tomonaga--Luttinger-liquid type) transition between the canted Ising and Haldane-like regimes, rather than a conventional Ising universality class. A complementary finite-size analysis at larger anisotropy (Fig.~\ref{scale_S_hx2c0}) demonstrates that the field range associated with the Haldane-like regime (low-field maximum) shrinks rapidly as $\alpha$ increases. Furthermore, the second maximum associated with the finite-size transition between phases III and VI shifts toward lower fields and gradually merges with the low-field maximum as the system size increases, suggesting a common singular point at $h_x=0$ in the thermodynamic limit. Consequently, while Haldane-like signatures remain robust close to the isotropic limit, their stability is progressively reduced away from $\alpha=1$, indicating that this regime is primarily tied to the vicinity of the isotropic Heisenberg point. For this reason, we refer to phase III as a Haldane-like regime rather than a genuine Haldane phase, since the available finite-size data support the persistence of Haldane characteristics in a narrow neighborhood of $\alpha=1$, while their precise thermodynamic extent remains unresolved.

\subsubsection{Luttinger liquid phase}

The rung spin–spin correlations at $\alpha = 1$ are nearly isotropic, with $C_{1,N/4}^{\gamma} \equiv \tfrac{1}{3} C_{\rm rung} = +\tfrac{1}{12}$ for each spin component $\gamma = x,y,z$, indicating the formation of effective rung-triplet states and supporting an emergent spin-1 description of the ladder. The corresponding bulk value $C_{\rm rung} = +\tfrac{1}{4}$ reflects strong ferromagnetic alignment within each rung in the effective representation. In contrast, the leg correlations remain finite and antiferromagnetic, with $C_{\rm leg} = a \approx -0.350371$, corresponding to $C_{2,N/4}^{\gamma} \equiv \tfrac{1}{3} C_{\rm leg} \simeq -0.117$ for each component $\gamma = x,y,z$, indicating residual inter-rung quantum fluctuations consistent with a weakly coupled spin-1 chain picture.

\begin{figure}[t!]
\includegraphics[width=1\columnwidth]{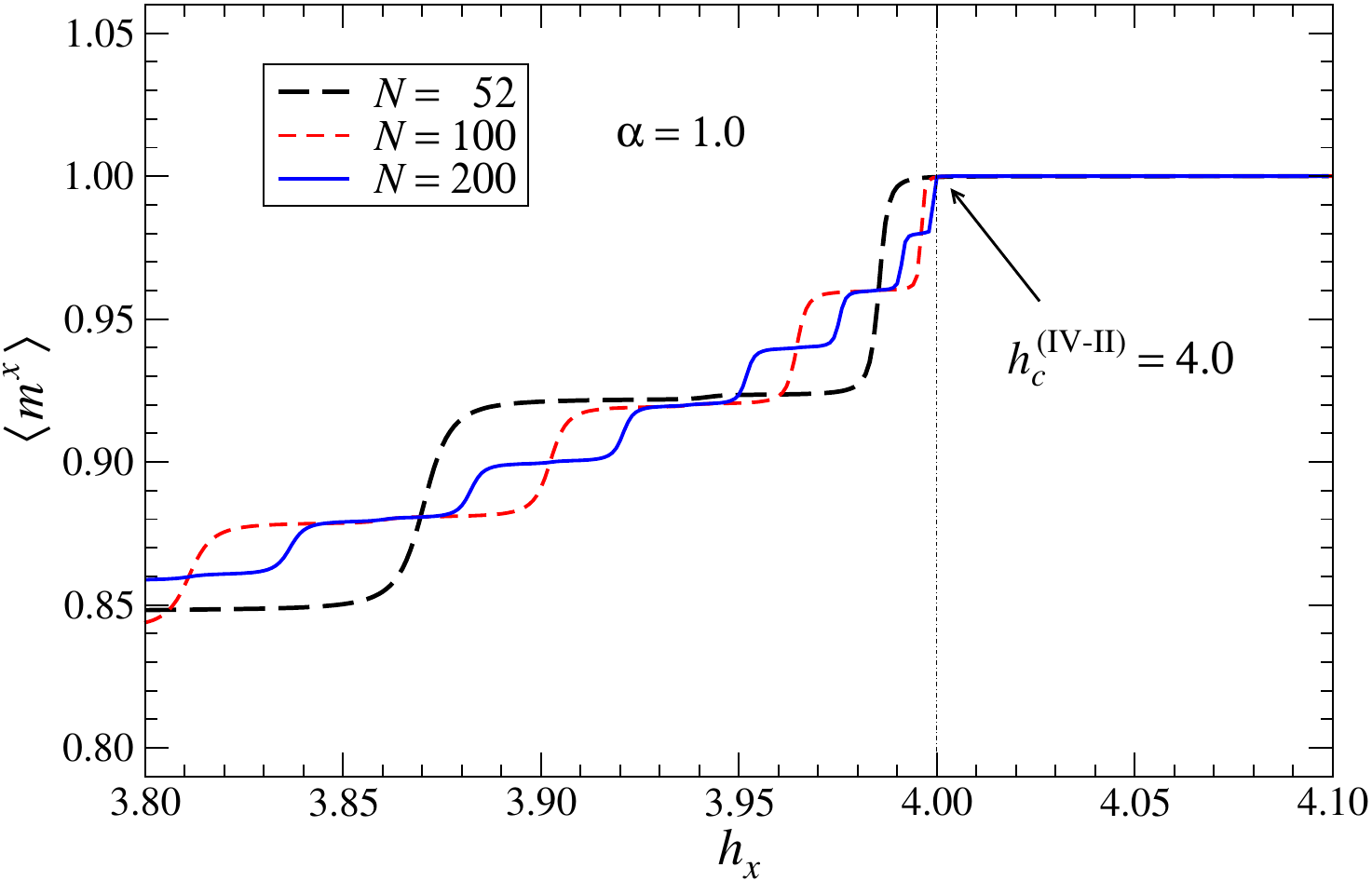}
  \caption{Size dependence of the staircase structure in the rung magnetization $\langle m^x \rangle$ at the central rung within the Tomonaga–Luttinger liquid phase IV. Results are shown for $\alpha = 1$ and system sizes $N = 52$, $100$, and $200$. The vertical dot-dashed line marks the transition at $h_c^{\rm (IV-II)} = 4$. 
  }
\label{fig:stair}
\end{figure}
The Tomonaga–Luttinger liquid phase, denoted as phase IV in Fig.~\ref{fig:fd}, is realized in the vicinity of the isotropic point $\alpha = 1$, where it forms a narrow region in the finite-size phase diagram. At $\alpha = 1$, the phase extends over the interval $0.44 \lesssim h_x \leq 4$, in agreement with previous results for the isotropic frustrated antiferromagnetic Heisenberg ladder~\cite{Honecker}.

\begin{figure}[t!]
{\includegraphics[width=1.0\columnwidth]{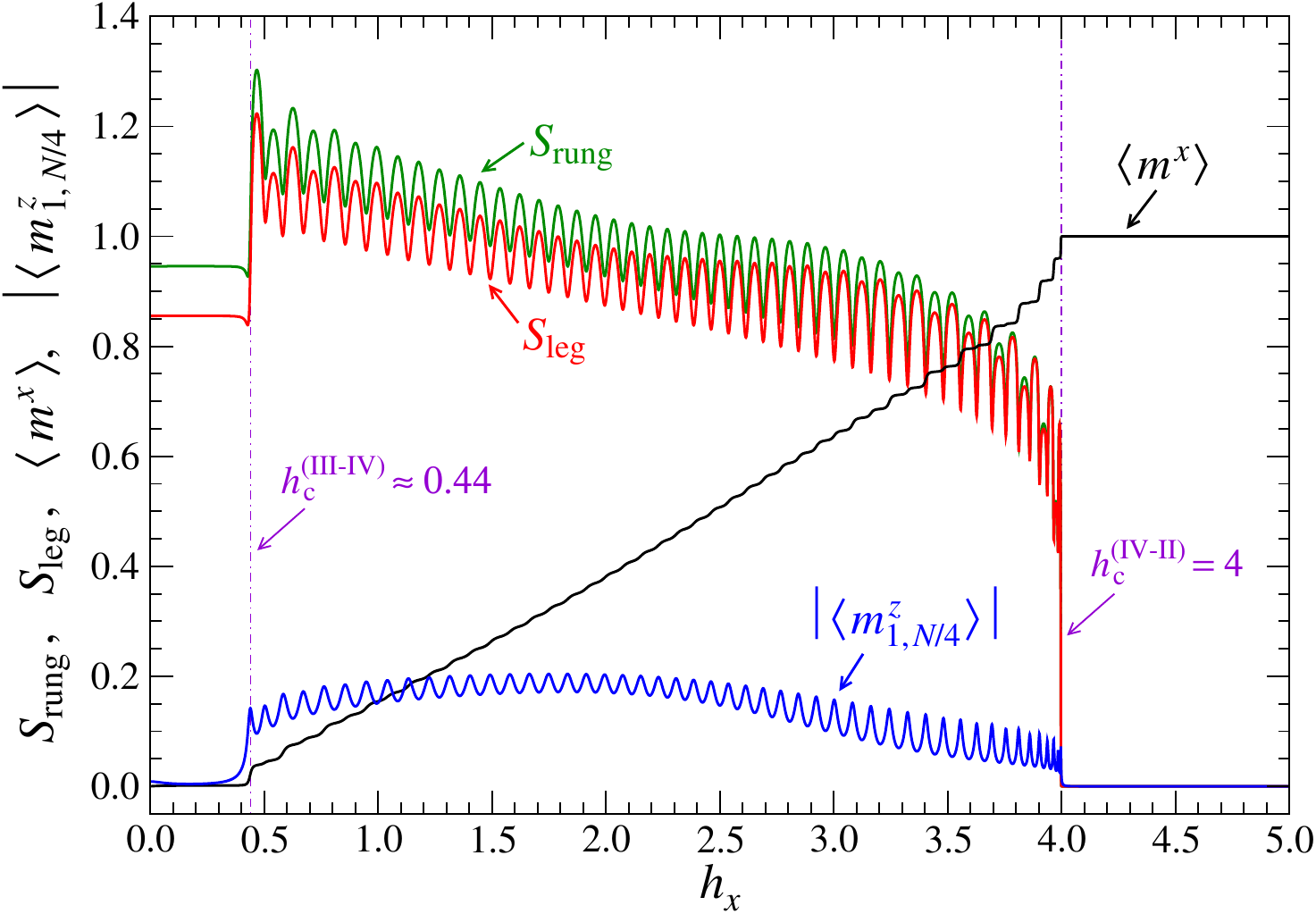}}
  \caption{Transverse-field dependence of the entanglement entropies $S_{\rm rung}$ and $S_{\rm leg}$, the rung magnetization at the central rung $\langle m^x \rangle$, and the magnitude of the onsite magnetization at the central site $|\langle m^z_{1, N/4} \rangle|$, calculated for $\alpha = 1$ and $N = 100$. Vertical dot-dashed lines indicate the phase transitions at $h_c^{\rm (III-IV)} \approx 0.44$ and $h_c^{\rm (IV-II)} = 4$.}
\label{fig:lutting}
\end{figure}

In the thermodynamic limit, the net magnetization $\langle m^x \rangle$ increases smoothly with the transverse field $h_x$, consistent with the absence of a magnetization gap in this phase. For finite ladders, $\langle m^x \rangle$ develops a staircase-like structure due to discrete changes in the total spin, which gradually smooths as the system size increases (Fig.~\ref{fig:stair}). Similar finite-size oscillations appear in the onsite longitudinal magnetization at a central site $|\langle m^z_{1,N/4}\rangle|$, reflecting boundary-induced effects that vanish in the thermodynamic limit (Fig.~\ref{fig:lutting}, blue curve). The local transverse magnetization $\langle m^x_{\ell,j} \rangle$, shown in Fig.~\ref{fig:edge}, exhibits  Friedel oscillations, a hallmark of one-dimensional criticality~\cite{Hikihara}, superimposed on a finite uniform background. For $N=100$, the latter reaches $\langle m^x_{1,N/4}\rangle \approx 0.077$, as illustrated in the inset of Fig.~\ref{fig:edge}.  The amplitude of these oscillations decays algebraically from the boundaries, reflecting the gapless, power-law correlations of the Tomonaga–Luttinger liquid. Unlike the topologically protected edge states of the Haldane phase, the enhanced edge magnetization here arises purely from critical boundary effects and does not indicate a gapped topological order.

The rung and leg entanglement entropies, $S_{\rm rung}$ and $S_{\rm leg}$, are nearly identical throughout the phase and exceed $\ln 2$, reflecting strongly entangled multipartite correlations. For finite ladders, both quantities exhibit field-induced oscillations across the parameter range (Fig.~\ref{fig:lutting}), which diminish with increasing system size and converge toward a smooth profile in the thermodynamic limit. The spatial profile displays a characteristic inverted-U shape, arising from open boundaries and boundary-induced effects, and is consistent with gapless critical correlations of the Tomonaga–Luttinger liquid rather than topological edge states (Fig.~\ref{fig:edge}, $h_x=1$, $\alpha=1$).

Rung correlations remain constant, $C_{\rm rung}= +\tfrac{1}{4}$, indicating robust ferromagnetic alignment within each rung. The dominant contribution evolves from $z$- to $x$-components with increasing $h_x$, reflecting the progressive field-induced rotation of the local spin quantization axis. Leg correlations evolve continuously from $C_{\rm leg} \approx -0.350371$ at low fields to $C_{\rm leg} \approx +\tfrac{1}{4}$, exhibiting strong anisotropy and sensitivity to both $\alpha$ and $h_x$: larger $\alpha$ stabilizes antiferromagnetic correlations along the legs, while higher $h_x$ drives a field-induced reorientation from $z$- to $x$-dominated correlations, consistent with the smoothly evolving correlations expected in the Tomonaga–Luttinger liquid regime.

The stability of the Tomonaga–Luttinger liquid phase is further supported by finite-size scaling of the rung entanglement entropy $S_{\rm rung}$ in the vicinity of $\alpha = 1$ (right panel of Fig.~\ref{scale_S_hx2c0}). As the system size increases up to $N=500$, the entropy exhibits a systematic growth and the finite interval of stability around $\alpha = 1$ progressively narrows, consistent with a collapse of the gapless regime to a fine-tuned line in the thermodynamic limit. In this regime, we further extract the central charge from the scaling of $S_{\rm rung}$ at $\alpha=1$ and $h_x=2$, obtaining $c \simeq 1$ within numerical accuracy, in agreement with the expected Tomonaga–Luttinger liquid universality class (see right panel in Fig.~\ref{CentralCarge}).

\subsubsection{Canted Ising-order phase} 

\begin{figure*}[!tb]
{\centering\includegraphics[width=\columnwidth]{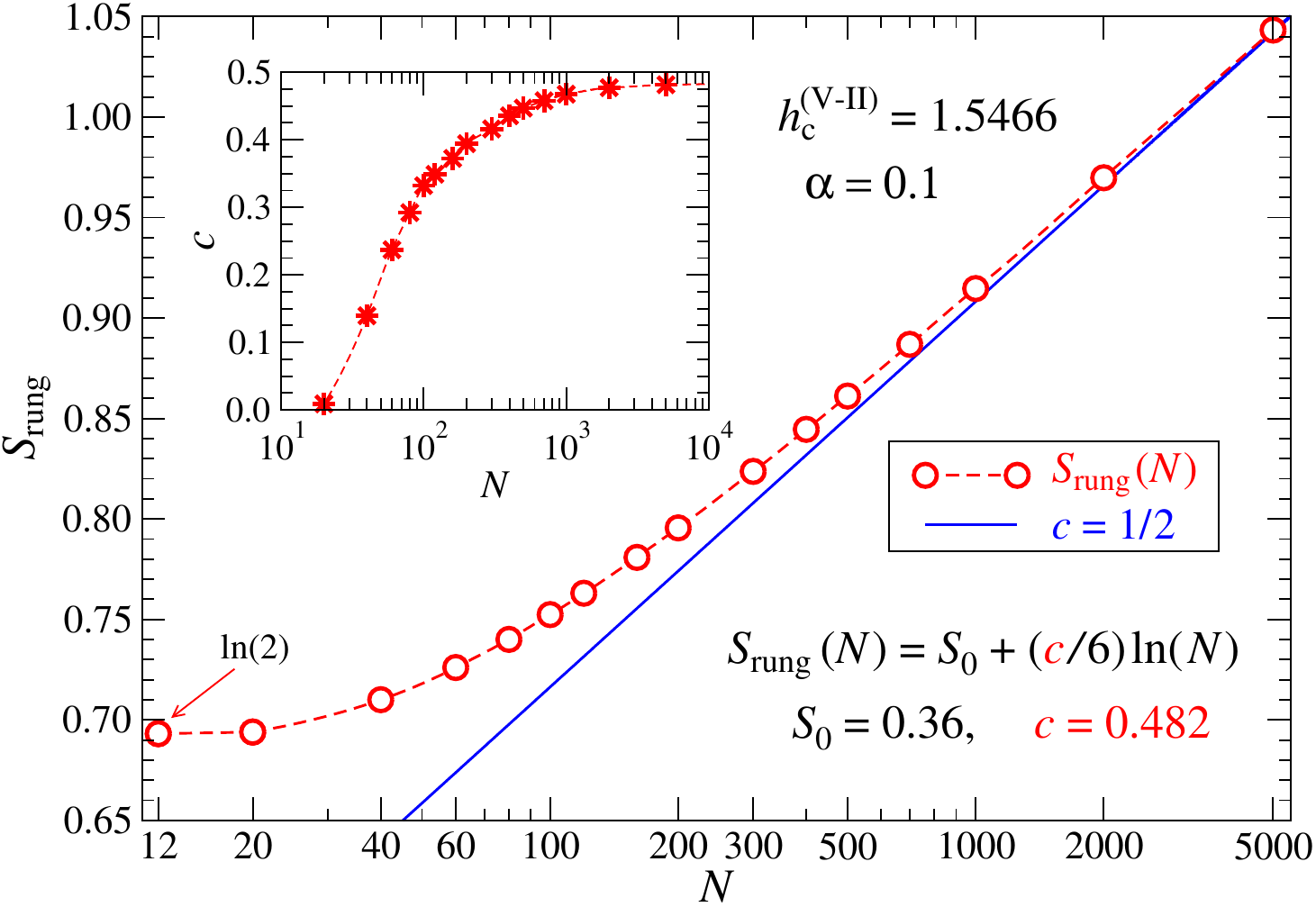}}\quad
{\centering\includegraphics[width=\columnwidth]{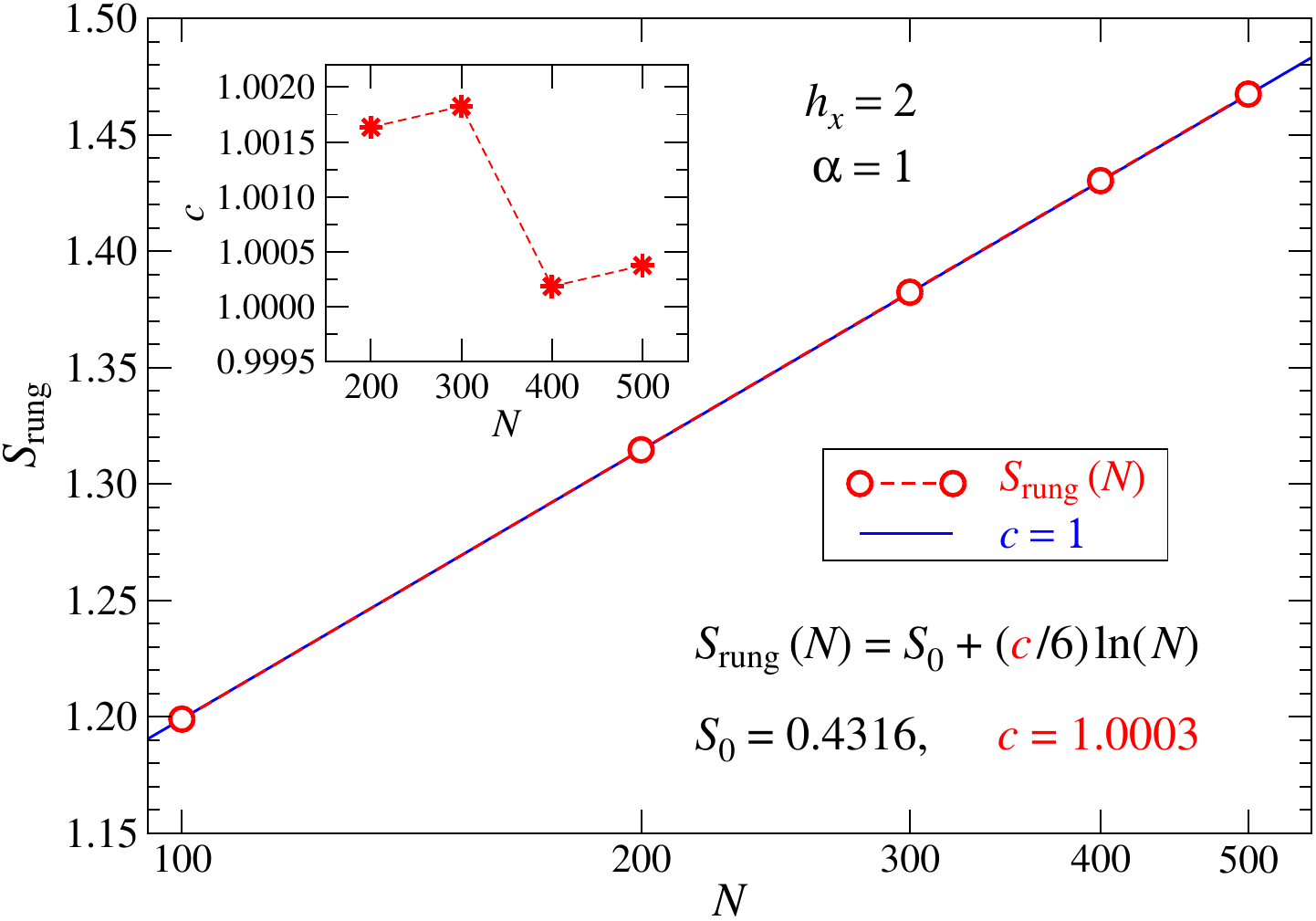}}
  \caption{Central-charge analysis from the asymptotic scaling of the rung entanglement entropy $S_{\rm rung}$ with system size $N$. Left panel: illustrates the transition between the canted Ising-ordered phase ${\rm V}$ and the ordered ferromagnetic phase ${\rm II}$ at $\alpha=0.1$ and $h_{\rm c}^{(\rm V-II)}=1.5466$ up to $N=5000$. Right panel: illustrates the behavior at $\alpha=1$ and $h_x=2$ within the Luttinger-liquid phase ${\rm IV}$ up to $N=500$. Blue lines in both panels denote the asymptotic scaling with central charge $c=1/2$ (left panel) and $c=1$ (right panel). Insets: finite-size extrapolation of the central charge obtained from the numerical derivative $c=6\,{\rm d}S/{\rm d}\ln N$.}
\label{CentralCarge}
\end{figure*}

The canted Ising phase, denoted as phase V in Fig.~\ref{fig:fd}, exists for $\alpha<1$ in the region separating phases I, II, and III. The net transverse magnetization $\langle m^x \rangle$ increases smoothly with the applied transverse field $h_x$ while the net longitudinal magnetization remains zero, $\langle m^z \rangle = 0$, owing to the staggered antiferromagnetic arrangement along both legs. This phase is distinguished by a finite onsite longitudinal magnetization, ferromagnetically ordered on each rung, $\langle m^z_{1,j} \rangle = \langle m^z_{2,j} \rangle$. The coexistence of these longitudinal and transverse components reflects a canted spin configuration, in which local moments are tilted away from the Ising axis toward the field direction. As illustrated in the right panel of Fig.~\ref{fig:mag}, this component gradually decreases from its saturated value of $\tfrac{1}{2}$ with increasing $h_x$ or anisotropy $\alpha$, eventually vanishing in the strong-field or large-$\alpha$ limit.

The rung and leg entanglement entropies, $S_{\rm rung}$ and $S_{\rm leg}$, are finite, nearly equal, and largely uniform across the lattice, remaining below $\ln 2$. Their magnitude grows with $\alpha$, indicating enhanced quantum fluctuations as isotropy is approached; nonetheless, entanglement remains moderate throughout the phase, consistent with a partially ordered yet correlated ground state.

Rung correlations satisfy $C_{\rm rung} = +\tfrac{1}{4}$, confirming ferromagnetic alignment within each rung. At low fields, longitudinal correlations $C_{\ell=1,j}^{z}$ dominate, signaling robust Ising-like behavior. With increasing $h_x$, spectral weight gradually shifts to the transverse correlations $C_{\ell=1,j}^{x}$, reflecting the continuous canting of spins toward the $x$-direction.

The critical behavior at the transition to the fully polarized phase is characterized by extracting the central charge $c$ from the finite-size scaling of the rung entanglement entropy at the critical field $h_{\rm c}^{\rm (V-II)}$. For $\alpha = 0.1$, calculations on lattices up to $N = 5000$ yield $c \approx 0.482$ (see left panel of Fig.~\ref{CentralCarge}), approaching $c = \tfrac{1}{2}$ in the thermodynamic limit. This value is consistent with the Ising universality class. 

\subsubsection{XY-polarized phase}

The XY-polarized phase, denoted as phase VI in Fig.~\ref{fig:fd}, occupies the region $\alpha > 1$ bounded by the phase-transitions into the ordered ferromagnetic boundary, i.e., $\vert h_c^{(\mathrm{VI-II})} \vert\leq 3\alpha + 1$. The net transverse magnetization $\langle m^x \rangle$ increases continuously with the applied field $h_x$ in this regime, starting from $\langle m^x \rangle=0$ at $h_x \to 0$ toward its saturation ($\langle m^x \rangle=1$) near the transition to the ordered ferromagnetic phase. In contrast to the canted Ising phase, both the net and onsite longitudinal magnetizations vanish, $\langle m^z \rangle = \langle m^z_{\ell,j} \rangle = 0$, indicating the absence of any longitudinal ordering.

The rung and leg entanglement entropies, $S_{\rm rung}$ and $S_{\rm leg}$, remain finite and saturate with system size, confirming a noncritical ground state throughout this phase, except for the phase-transition boundaries. Both quantities depend sensitively on the transverse field: for intermediate anisotropies, $S_{\rm rung}$ exhibits a broad maximum at intermediate fields and decreases upon approaching the ferromagnetic boundary, reflecting the progressive suppression of quantum correlations under strong polarization. For larger $\alpha$, this field dependence becomes weaker, consistent with a more rigid, interaction-dominated regime.

Spin–spin correlations are predominantly confined to the transverse plane, with $C^{y}_{\ell,j}$ providing the dominant contribution, while $C^{x}_{\ell,j}$ remains finite and subleading to $C^{y}_{\ell,j}$. The nearest-neighbor correlation $C^{z}_{\ell,j}$ is strongly suppressed to zero. For $\alpha \gg 1$, the relative weight of $C^{x}_{\ell,j}$ increases, indicating a gradual tendency toward XY isotropy in the thermodynamic limit. In contrast to the Ising-dominated regime at $\alpha < 1$, no staggered or symmetry-breaking longitudinal structure emerges; instead, the transverse correlations evolve smoothly with $h_x$, reflecting a continuous redistribution of quantum fluctuations within the XY sector driven by inter-rung exchange.

\section{Conclusions} \label{conclusion}

We have investigated the ground-state phase diagram of a frustrated spin-1/2 Heisenberg ladder in the transverse magnetic field $h_x$ with bond-dependent exchange anisotropy $\alpha$, which interpolates between the Ising model $(\alpha=0)$, the isotropic Heisenberg model $(\alpha=1)$, and the XY model $(\alpha \gg 1)$. Our results show that  the parameter $\alpha$ primarily governs the competition between quantum fluctuations and exchange anisotropy, thereby determining the global structure of the phase diagram. This establishes a unified framework connecting several distinct classes of quantum states within a single frustrated ladder system.

The system includes six distinct ground-state phases: the rung-singlet phase, the Haldane-like phase, the Tomonaga–Luttinger liquid, the canted Ising-ordered phase, the XY-polarized phase, and the ordered ferromagnetic phase. The resulting phase diagram exhibits a pronounced concentration of quantum-critical behavior near the isotropic point $\alpha = 1$, where  features of the Haldane-like phase and the Tomonaga–Luttinger liquid on a finite interval are stabilized. In contrast, deviations from isotropy rapidly suppress criticality and favor either a gapped or a field-polarized phase, demonstrating the remarkable fragility of critical quantum states to exchange anisotropy in frustrated low-dimensional magnets.

To characterize these phases on equal footing, we employ bipartite entanglement and its finite-size scaling. The extracted central charge $c$ distinguishes the underlying universality classes: we obtain $c \simeq 1$ in the Tomonaga–Luttinger liquid regime, consistent with gapless critical behavior, and $c \simeq 1/2$ at the canted Ising–ferromagnetic transition, in agreement with Ising universality. In contrast, all gapped or field-polarized phases are characterized by saturated entanglement entropy and vanishing central charge in the thermodynamic limit ($N\to\infty$).

Beyond the conventional characterization based on magnetization and local correlation functions, our results demonstrate that bipartite entanglement provides direct access to the underlying structure of quantum correlations in frustrated ladder systems. In particular, the finite-size scaling of entanglement entropy enables a clear distinction between critical and noncritical regimes even in parameter regions where local magnetic observables exhibit qualitatively similar behavior. This becomes especially important near the isotropic limit, where competing phases exhibit similar magnetic signatures despite fundamentally different correlation properties and universality classes. Our results therefore demonstrate that entanglement-based diagnostics is not merely a complementary tool but may become essential for resolving competing quantum phases in frustrated systems with strong correlations and reduced dimensionality. The present findings thus provide broader insight into the interplay between frustration, anisotropy, and quantum criticality in low-dimensional quantum magnets.

\begin{acknowledgments}
   This work was supported by the Slovak Research and Development Agency under the Contract No. APVV-24-0091, by the Scientific Grant Agency of the Ministry of Education, Science, Research and Youth of the Slovak Republic and the Slovak Academy of Sciences through grants No. VEGA 2/0152/26 and No. VEGA 1/0298/25, by the
   Stefan Schwarz fund No. 2024/OV1/026, and by the EU NextGenerationEU through the Recovery and Resilience Plan for Slovakia under the project No. 09I03-03-V04-00682.
\end{acknowledgments}

\end{document}